\input xy
\xyoption{all}

\documentclass[12pt]{article}
\input epsf.tex
\usepackage{pstricks}
\usepackage{psfrag}
\usepackage{graphicx}
\usepackage[small,loose]{subfigure}
\usepackage[latin2]{inputenc}
\usepackage{epic}
\usepackage{latexsym}
\usepackage{mathrsfs}
\usepackage{bbm} 
\usepackage{cancel} 
\usepackage{amssymb,amsmath}
\usepackage{fleqn} 

\def\harr#1#2{\smash{\mathop{\hbox to .3in{\rightarrowfill}}
 \limits^{\scriptstyle#1}_{\scriptstyle#2}}}

\def\yzero{\smash{\hbox{$y\kern-4pt\raise1pt\hbox{${}^\circ$}$}}}

\def\sin2{\frac{1}{\sqrt2}}

\def\be{\begin{equation}}
\def\ee{\end{equation}}
\def\beqa{\begin{eqnarray}}
\def\eeqa{\end{eqnarray}}
\def\bl{\begin{align}}
\def\el{\end{align}}

\def\Dsl{\,\raise.15ex\hbox{/}\mkern-13.5mu D} 

\def\bfr{\begin{flushright}}
\def\efr{\end{flushright}}

\def\IZ{\mathbb{Z}}

\def\N{{\cal N}}
\def\W{\mathscr{W}}
\def\Nf{{\cal N}_{\text{$D3$/flux}}}

\def\IRP{\ensuremath{\mathbb{R}{\rm P}}}

\def\*{\ast_{(6+1)}}

\topmargin
-.5cm
\textwidth
15.5cm
\textheight
23cm
\oddsidemargin
0.7cm
\evensidemargin
1.2cm

\begin{document}
\makeatletter
\@addtoreset{equation}{section}
\makeatother
\renewcommand{\theequation}{\thesection.\arabic{equation}}


\vspace{.5cm}
\begin{center}
\Large{\bf Minkowski vacuum transitions in (non-geometric) flux compactifications}\\

\vspace{1cm}

\large
Wilberth Herrera-Su\'arez\footnote{e-mail :{\tt wherrs@fis.cinvestav.mx}} and
Oscar Loaiza-Brito\footnote{e-mail :{\tt oloaiza@fis.cinvestav.mx}}\\[2mm] 

{\small \em Centro de Investigaci\'on y de Estudios Avanzados del I.P.N., Unidad Monterrey}\\
{\small\em Autopista Monterrey-Aeropuerto Km 10, 66600 Apodaca, Nuevo Le\'on, M\'exico}\\[4mm]

\vspace*{2cm}
\small{\bf Abstract} \\
\end{center}

\begin{center} 
\begin{minipage}[h]{14.0cm} { 
In this work
we study the generalization of twisted homology to geometric and non-geometric backgrounds.  In the process we describe the necessary conditions to wrap a network of D-branes on twisted cycles. If the cycle is localized in time,  we show how by an instantonic brane mediation, some D-branes transform into fluxes on different backgrounds including non-geometric fluxes. As a consequence, we show that in the case of a IIB six-dimensional torus compactification on a simple orientifold, the flux superpotential is not invariant by this brane-flux transition, allowing the connection among different  Minkowski vacuum solutions. 
For the case in which non-geometric fluxes are turned on, we also discuss some topological restrictions  for the transition to occur.  In this context, we show that there are some vacuum solutions protected to change by a brane-flux transition.
}
\end{minipage} 
\end{center}

\bigskip

\bigskip
 
\bigskip
 


\newpage


\section{Introduction}

Incorporation of fluxes in string theory compactification has been a rich and fertile topic over the last decade \cite{Grana:2005jc, Douglas:2006es}. Besides the important possibility to stabilize (perturbatively) all moduli at the minimum of a scalar potential generated by the fluxes, the presence of new kinds of fluxes has enriched our comprehension about geometry and topology on the internal spaces on which string theory is compactified.\\

The traditional way to realize the existence of these new fluxes, comes from performing T-duality on a type IIB compactification threaded with NS-NS flux. In this way,  it is necessary to introduce fluxes referred to as {\it metric and non-geometric}.  
In \cite{Shelton:2005cf} (see also \cite{Wecht:2007wu} for a review) was shown that the inclusion of non-geometric fluxes\footnote{For recent advances in non-geometric flux compactifications, see \cite{Aldazabal:2008zza, Font:2008vd, Guarino:2008ik}.} in a tori compactification leads to two important facts. First, that the metric of a torus threaded with non-geometric fluxes  ($Q$-space) is not globally defined, and second, that the corresponding IIB  superpotential does not preserve S-duality.\\

The first statement means that it is impossible to consistently define one-cycles on the 6-dimensional corresponding torus on which some non-geometric flux is supported. 
The reason for this  comes from the application of Buscher rules to the flat torus metric, from which one gets that the metric is not invariant after performing a translation along  one coordinate. Hence, the corresponding 1-cycle is also ill-defined.\\

However, if one T-dualizes  back to the $H$-space, the problem translates into a configuration involving  NS-NS flux $H_3$.  Specifically, it is inconsistent to define three-cycles on which the NS-NS-form is supported.  This is the classical version of  Freed-Witten anomaly \cite{Freed:1999vc}. Traditionally, in the literature the anomaly is cancelled by  enforcing the pull back of fluxes to the D-branes to vanish.  An alternative way to cancel it is to add extra D-branes. \\

In this context, a NS-NS flux pullbacks on the worldvolume of a D3-brane inducing the term $H_3\wedge A_1$ over the four dimensional worldvolume. The equations of motion for the  gauge field $A_1$ are fulfilled once we add magnetic sources, otherwise the system suffers of an anomaly.
By adding extra D1-branes ending at the former, the anomaly is  cured. T-dualizing forward in the appropriate coordinates, leads us to a configuration involving a net of D-branes in which is possible to wrap D1-branes in any one-cycle of the Q-space as far as we consider D3-branes 
with one leg along a non-compact direction and the other two wrapping internal coordinates, 
implying that we also require to twist the notion of a form.
Notice that in the $H$-space, we can safely take zero-, one- and two-cycles globally, but bigger ones require extra branes if NS-NS-flux is present. The same applies for the $Q$-space, but in this case we can only safely consider zero-cycles. \\

Yet, there are some D-branes that despite of being non anomalous,  could be unstable to decay into fluxes.  In this paper, we shall concentrate in the above way to cancel the anomaly which does not require the fluxes to vanish at the D-branes. In terms of forms, the Freed-Witten anomaly implies the existence of an obstruction to define a 3-form supported on the same coordinates as $H_3$. This makes sense as soon as we think on a cycle as the locus on which we can safely wrap a D-brane \cite{Maldacena:2001xj}. \\

Hence, by considering an adequate set of submanifolds (chains), we can wrap D-branes on generalized cycles consisting on $p$ and $(p-2)$-chains. An appropriate classification of such network is given in terms of twisted homology in \cite{Evslin:2007ti, Collinucci:2006ug}. Our first purpose is then, to extend the notion of twisted homology to the case of non-geometric fluxes. In this formalism, we will be able to elucidate the correct generalized cycle required to wrap D-branes in the presence of non-geometric fluxes.\\

Along  the way, we face the interesting case in which the anomalous brane is localized in time, establishing a (topological) mechanism by which  some D-branes decay into fluxes in the presence of background fluxes. Our results explicitly show how non-geometric fluxes drive some D-branes to transform into a coupling of RR fluxes (attributed to the instantonic brane) and the already present non-geometric fluxes. We find that a D$p$-brane would decay into fluxes by encountering an instantonic $E(p-2)$-brane supporting non-geometric flux. Thus, the impossibility to wrap D-branes on certain one-cycles is overcome by adding extra D-branes.\\

However, the transition between branes and fluxes has another consequence. As a result of the appearance of an instantonic brane, one extra unit of RR magnetic field strength adds to the flux background. Although the total amount of D-brane charge, coming from fluxes and physical D-branes, is constant, relevant functions depending explicitly on the units of RR flux would change.  An example of such function, is the Gukov-Witten-Vafa flux superpotential $\W$  which is not invariant under the cancelation of Freed-Witten anomalies by adding extra branes. In consequence, the Minkowski vacuum solutions gathered from it will be connected via the brane-flux transition\\

In the second part of this paper, we focus then to the study of the effects of brane-flux transitions on the Minkowski vacuum solutions. We consider the cases on which there are only NS-NS  and non-geometric fluxes and their S-dual versions which address the problem mentioned at the beginning. Interesting enough, we find  connections between different vacuum solutions, suggesting  a shortening on the landscape of solutions. Also, we find some cases in which the solution is protected to change under the brane-flux transitions, suggesting that there are places in the landscape which are isolated from others via this mechanism.\\

Our paper is organized as follows.  In section 2 we briefly review the concept of twisted homology in the presence of NS-NS flux. Later on, we extend this concept to the case in which the compact space is threaded with non-geometric fluxes. For completeness we also consider the case of geometric fluxes.  We study in detail the brane configuration and the required twisted cycles which give rise to the transition between D-branes and fluxes mediated by the appearance of instantonic branes.\\

In section 3 we focus on the effects of such transitions on the Minkowski vacuum solutions, supersymmetric and non-supersymmetric, on a type IIB compactification on a torus in  a simple orientifold ($O3^-$) threaded with 3-form fluxes while in section 4 we study the case with non-geometric fluxes.
We also analyze the required conditions under which the Bianchi identities are satisfied before and after the transition. This  establishes some constraints to be fulfilled by the configuration of fluxes for the transition to happen.  Since a general description turns out to be complicated, we have focused on some concrete examples. Finally we give our conclusions and some comments in section 5. The appendix  A is devoted to those interested on the cap product, extensively used along this paper to construct the required twisted cycles.\\

Finally, we would like  to remark that the transition between branes and fluxes would occur only if the transition is energetically favorable. Since our description concentrates just on topological arguments, such question will not be addressed in this paper.\\

\section{Twisted homology}

In this section we shall show an appropriate  formalism to describe consistent configurations of branes and fluxes,  focussing  on the familiar case of a tori compactification with NS-NS and RR 3-form fluxes\footnote{Strictly speaking, a compactification on a torus threaded with 3-form fluxes is not a solution of supergravity equations. A correct background involving the presence of an orientifold 3-plane will be considered later on.}. In the process, we shall describe a particular configuration involving a D-brane localized in time, namely an instantonic D-brane, by means of which some D-branes are unstable to decay into fluxes \cite{Maldacena:2001xj}.

\subsection{$H$-Twisted homology}
Let us start by reviewing how a cycle is defined in terms of simplexes, and how such cycles are not closed once an NS-NS-flux is turned on. We closely follow  \cite{Collinucci:2006ug} where a definition of twisted homology is given.\\

A $p$-cycle on a compact manifold $M$ is defined as
$[\Pi_p]=\sum_{j}C^j(\sigma_{p})_j$,
where $(\sigma_p)_j$ is the $j$-th $p$-simplex and $C^j$ the corresponding integer associated to $(\sigma_p)_j$ such that $[\Pi_p]$ is boundary-less and it is not the boundary of a lower dimensional chain. By an abuse of notation we shall denote the $p$-cycle simply as $[\Pi_p]=C^{\mu_1\cdots\mu_p}\sigma_{\mu_1\cdots\mu_p}$.  Integer homology is enough to correctly classify such submanifolds. The corresponding co-cycles are classified by cohomology. However if one turns on NS-NS 3-form fluxes, the supergravity equations of motion twist the co-boundary on the co-cycles. This is, instead of having closed RR forms, we have that
$dF_p=H_3\wedge F_{p+2}$.\\

As shown in \cite{Collinucci:2006ug}, one can instead define a twisted differential operator $d_H=d+H_3\wedge$ acting on poly-forms, such that they are closed under $d_H$. Similarly, one can twist the notion of cycle such that they are boundary-less in the presence of NS-NS 3-form fluxes. We proceed to describe the corresponding twisted homology studied in \cite{Collinucci:2006ug}.\\

Let $\partial_H$ be defined as\footnote{See Appendix A for some useful definitios as the cap product $\cap$ and its applications.}
$\partial_H=\partial + H_3\cap$.
An $H$-twisted cycle  ($H$-cycle for short) $[\Pi]$ is a network of chains\footnote{Throughout this paper, we shall use the square brackets $[~]$ to denote the homology class defined by a boundary-less chain.}  on which we can wrap D-branes such that the brane configuration is anomaly free. Consider for instance a D$p$-brane wrapping a cycle $[\Pi_p]$  where $[\Pi_p]\in \text{H}_p(M;\IZ)$. There are also $N$ units of NS-NS flux $H_3$ supported on a 3-cycle $[\Sigma_3]\in [\Pi_p]$. Notice that $[\Pi_p]$ is not an $H$-cycle since
\begin{align}
\partial_H [\Pi_p]= H_3\cap [\Pi_p]= N [\Pi_{p-3}],
\end{align}
where $[\Pi_{p-3}]$ is the co-dimension 3 cycle in $[\Pi_p]$ where the $H_3$-induced monopole should lie.  However if one adds a D$(p-2)$-brane supported on a chain  $\Sigma_{p-2}$ which ends at the D$p$-brane, i.e., $\partial \Sigma_{p-2}= [\Pi_{p-3}]$, hence we get that the poli-cycle $[\Pi]=[\Pi_p]-N\Sigma_{p-2}$ is an $H$-cycle since
\begin{align}
\partial_H[\Pi]= H_3\cap [\Pi_p] - N\partial \Sigma_{p-2}= 0.
\end{align}
This tells us that in order to have an anomaly free configuration of branes and fluxes, a D$(p-2)$-brane should end at a D$p$-brane supporting one unit of NS-NS-flux. For $N$ units, we require the presence of $N$ D$(p-2)$-branes.\\

Notice however that in a compact space, the lower dimensional brane wraping  an internal  cycle,  intersects the one on which the D$p$-brane is wrapped. One can think of this as if the lower dimensional brane has two boundaries immersed at the cycle on which the NS-NS-flux is supported. In this context, the anomaly is not cured.\\

\noindent
{\bf Example.}
Take  a six-dimensional torus $T^6$ threaded with an NS-NS-flux $H_3=N dx^1\wedge dx^2\wedge dx^3$, where coordinates $(x^i, x^{i+3})$ with $i=1,2,3$ parametrize the ith-torus $T^2_i$. Consider as well the basis for one-cycles denoted as $([\sigma_i], [\sigma_{i+3}])$ such that
$\int_{[\sigma_i]}dx^j=\int_{[\sigma_{i+3}]}dx^{j+3}=\delta^j_i$ and 
$\int_{[\sigma_i]}dx^{j+3}=\int_{[\sigma_{j+3}]}dx^j=0$.
Take now a D3-brane wrapping the cycle
\begin{align}
[\Pi_3]=\bigotimes_{i=1}^3 n^i[\sigma_i]=n^{123}\sigma_{123}.
\end{align}
The cycle $[\Pi_3]$ is clearly not an $H$-cycle since $\partial[\Pi_3]=0$ and
\begin{eqnarray}
H_3\cap [\Pi_3]&=& H_{123}n_1n_2n_3\left(dx^1\wedge dx^2\wedge dx^3\right)\cap \left([\sigma_1]\otimes[\sigma_2]\otimes[\sigma_3]\right)\nonumber\\
&=&H_{123}n^1n^2n^3[0]\nonumber\\
&=& Nn^1n^2n^3 [0],
\end{eqnarray}
where $[0]$ is the zero-cycle in $T^6$ generating $H^0(T^6, \IZ)$. Let us instead consider as well the presence of D1-branes wrapping for instance coordinate $x^4$.  The representing chain in this case is denoted as $\Sigma_1=b^4\sigma_4$ (and by construction $\partial \Sigma_1= [0]\subset [\Pi_3]$). Hence the poli-cycle given by
\begin{align}
[\Pi]= [\Pi_3]- Nn^1n^2n^3\Sigma_1,
\end{align}
is not an $H$-cycle either since, as mentioned, $[\Sigma_1]$ can be thought as  having two boundaries at $[\Pi_3]$,
\begin{align}
\partial_H[\Pi]= H_3\cap[\Pi_3]-Nn^1n^2n^3\partial \Sigma_1+Nn^1n^2n^3\partial\Sigma_1= Nn^1n^2n^3[0].
\end{align}
A truly $H$-cycle is given by a D1-brane which extends in transversal coordinates to $T^6$. For that to happen, let us consider a 10-dimensional space-time $M=X_4\times T^6$. In such a case, a D1-brane wraps a spatial direction in $X_4$ denoted as $\Sigma_x$ and sitting at a point in $T^6$ (i.e., $\partial \Sigma_x=[0]\in T^6$).  The poli-cycle $[\Pi]=[\Pi_3]-Nn^1n^2n^3\Sigma_x$ satisfies
$\partial_H [\Pi]=0$.\\

Notice that we can interpret this result as the fact that in the presence of $N$-units of flux supported on the worldvolume of  a D3-brane with warping numbers $n^i$, we require for consistency $Nn^1n^2n^3$ D1-branes ending at it. This is equivalent to say that we have a D3-brane with warping number equal to one and $Nn^1n^2n^3$ units of NS-NS flux.  

\subsubsection{Brane-flux transition}
Looking for configurations free of anomalies, or equivalently by classifying $H$-cycles, leads us to systems in which a D-brane transforms topologically into fluxes. This phenomena was studied in detail in \cite{Maldacena:2001xj} (see also \cite{Diaconescu:2000wy}). As a matter of an example, consider a D5-brane localized in time (an instantonic E5-brane) wrapping the  6-cycle conveniently written as
\begin{align}
[\Pi_6]=[\Pi_3]\otimes[\Sigma_{xyz}],
\end{align}
where
$[\Pi_3]=\otimes_{i=1}^3[\sigma_i]$ is an internal cycle in $T^6$ and
$[\Sigma_{xyz}]$ is the 3-dimensional cycle in the extended spacetime. In the presence of an NS-NS-flux given by $H_3= Ndx^1\wedge dx^2\wedge dx^3$
the poli-cycle $[\Pi]$ given by
\begin{align}
[\Pi]=[\Pi_6]- N\Sigma_4,
\end{align}
with $\Sigma_4=[\Sigma_{xyz}]\times (-\infty, t_0]$ is clearly and $H$-cycle since $\partial_H[\Pi]=0$.
Here, $t_0$ is the time-like coordinate at which the instantonic brane appears. Hence, we see that $N$ D3-branes would disappear by encountering the instantonic E5-brane. After $t_0$, there is a remnant of flux due to the instantonic E5-brane, actually a 3-form flux $F_3$ supported in $[\Sigma_3]=\otimes_{i=1}^3[\sigma_{i+3}]$. This is the way in which $N$-branes transform into a configuration of fluxes\footnote{In the present paper we have focused our study on the topology of such  transition. For the corresponding dynamics, one should consider the Myers' effect at the level of the DBI action. A very illustrative and important example is studied in \cite{Kachru:2002gs}.}  given by $H_3\wedge F_3$. Note that after $t_0$, there is an extra unit of RR-flux $F_3$.

The would-be unstable D-branes decaying into flux\footnote{These D-branes will actually decay into fluxes if the transition is energetically favorable.} wrap submanifolds of $T^6$ which are $N$-torsion classes in the corresponding
$H$-homology \cite{Collinucci:2006ug} defined as
\begin{align}
\text{H}_{H,\ast}(X,;\IZ)= \text{Ker } \partial_H/\text{Im }\partial_H.
\end{align}

\subsection{$f$-twisted homology}
At the level of cohomology, it was shown in \cite{Marchesano:2006ns} that unstable type II D-branes in presence of non-trivial NS-NS-flux $H_3$  T-dualize into D-branes wrapping $N$-torsion cycles in ordinary integer  homology (i.e., with no fluxes involved). The number $N$ is then related to the amount of units of NS-NS-flux $H_3$ in the T-dual picture. Specifically we see  (\cite{Kaloper:1999yr, Kachru:2002sk, Camara:2005dc}) that it is possible to define a basis of non-closed 1-forms $\eta^j$ as
\begin{align}
\eta^a=dx^a-\frac{1}{2}f^a_{bc}x^bdx^c,
\end{align}
such that
\begin{align}
d\eta^a=-\frac{1}{2}f^a_{bc}~\eta^b\wedge \eta^c,
\label{deta}
\end{align}
with the twisted torus metric given by $ds^2=\eta_{ij}\eta^i\eta^j$. Our purpose is to define the T-dual version of the previously studied twisted homology  . 
For that, let us consider the two-form,  defined by the structure constant $f^a_{bc}$,  given by
\begin{align}
f_{(a)}=-\frac{1}{2}f^a_{bc}~\eta^b\wedge \eta^c,
\end{align}
and called {\it metric fluxes}. In addition,  
define the action of the metric fluxes on a $p$-chain $\Pi_p=b^{\mu_1\cdots\mu_p}\sigma_{\mu_1\cdots\mu_p}$ by the cap product between it and the 2-form $f_{(a)}$ as,
\begin{eqnarray}
f\cdot\Pi_p=f_{(a)}\cap\Pi_{p(a)}&=&-\frac{1}{2}(f^a_{bc}\eta^{bc})\cap(b_a^{\mu_1\cdots\mu_{p+1}}\sigma_{\mu_1\cdots\mu_{p+1}})\nonumber\\
&=&-\frac{1}{2}f^a_{bc}b_a^{\mu_1\cdots b\cdots c\cdots\mu_{p+1}}\sigma_{\mu_1\cdots\hat{b}\cdots\hat{c}\cdots\mu_{p+1}}
\end{eqnarray}
where $\hat{b}$ and $\hat{c}$ are suppressed indexes and we have made use of the definition
\begin{align}
\Pi_{p(a)}=b_a^{\mu_1\cdots\mu_{p+1}}\sigma_{\mu_1\cdots\mu_{p+1}},
\end{align}
where the index $a$ is contained in the set $\{ \mu_1,\cdots\mu_{p+1}\}$. From this we can see that  $\Pi_{p(a)}$ is actually a $(p+1)$-chain. This is as follows.
By a redefinition of coefficients  $b_a^{\mu_1\cdots\mu_{p+1}}\sim C_a^{bc}\tilde{b}^{\mu_1\cdots \hat{b}\cdots \hat{c}\cdots\mu_{p+1}}$,
we get
\begin{align}
f_{(a)}\cap\Pi_{p(a)}=-\frac{1}{2}f^a_{bc}C^{bc}_a( \tilde{b}^{\nu_1\cdots\nu_{p-1}}\sigma_{\nu_1\cdots\nu_{p-1}}),
\end{align}
where we have rearranged the indexes in the $(p-1)$-simplexes. Note that the second part of the rhs in the above equation constitutes by itself a $(p-1)$-chain, such that one can write the action of $f$ on a $p$-chain as\footnote{One can always normalize the chain $\Pi_p$ such that $C^{bc}_a=(\delta^{bc})_a$.}
\begin{align}
f\cdot\Pi_p=-\frac{1}{2}f^a_{bc}C^{bc}_a\Sigma_{p-1}.
\end{align}
Thus, $f$ acts on a chain as the boundary operator, since it reduces by one the dimension of the chain \footnote{Notice that $f^a_{bc}$ must be an even number for a  well definition of $f\cdot$. This translates in the $H$-space as the abscence of subtleties with exotic orientifolds as mentioned \cite{Frey:2002hf}.}.  Actually, from (\ref{deta}) we see that
\begin{align}
\partial\Pi_p=-f\cdot\Pi_p=N\Sigma_{p-1}.
\end{align}
We see that indeed, $\Sigma_{p-1}$ is an $N$-torsion cycle as described in \cite{Marchesano:2006ns}. 
Then by defining the $f$-twisted boundary as
$\partial_f=\partial+f\cap$, one gets
\begin{align}
\partial_f[\Pi_p]=\partial[\Pi_p]+f_{(a)}\cap\Pi_{p(a)}=0.
\end{align}
 Notice as well that  the $p$-chain $\Pi_p$ can be decomposed as
\begin{align}
\Pi_p=[\Theta_p]+\Xi_p(x),
\end{align}
where $\partial[\Theta_p]=0$ and $\partial\Xi_p\neq 0$,  i.e.  $[\Theta_p]\in H_p(X_9;\IZ)$ but $\Xi_p$ is not a cycle.\\

\noindent
{\bf Example.} Consider a compactification of 10-dimensional space-time  on a 6-dimensional torus $T^6$ threaded with metric flux $f_{(1)}=-\frac{1}{2}f^1_{56}\eta^5\wedge\eta^6$. The action of the metric flux on the 2-cycle $\Pi_2=b^{56}\sigma_{56}$ is given by
\begin{align}
f\cdot\Pi_2=-\frac{1}{2}N\tilde{b}^1\sigma_1,
\end{align}
where the 1-cycle $[\Sigma_1]=\tilde{b}^1\sigma_1$ turns out to be the boundary of $\Pi_2$. Thus, $[\Sigma_1]$ is an $N$-torsion cycle. Written in terms of coordinates $x^i$,
\begin{align}
\Pi_2=[\Theta_2]+\Xi_2(x^6),
\end{align}
where $\Theta_2$ is the Poincar\'e dual of $dx^1\wedge dx^2\wedge dx^3\wedge dx^4$ and $\Xi_2(x^6)$ is the Poincar\'e dual of the non-closed form $f^1_{56}x^6dx^2\wedge dx^3\wedge dx^4\wedge dx^5$. Hence, since $[\Theta_2]=b^{56}\sigma_{56}(x)$ we get that
$f\cdot [\Theta_2]= N[\Sigma_1]$ and  $\partial[\Theta_2]=0$. Hence, in order to recover that $\partial_f[\Pi_2]=0$ we must have that
\begin{align}
\partial \Xi_2(x^6)=N[\Sigma_1].
\end{align}
This implies that a D2-brane wrapping the 2-cycle $[\Theta_2]$ in the presence of metric flux $f^1_{56}(=N)$ requires $N$ D2-branes wrapping the chain $\Xi_2$ which has a boundary contained in $[\Theta_2]$.\\

Some comments are in order: First, notice that there are no anomalies concerning the wrapping of D1-branes or D0-branes on one- and zero-cycles since $f\cdot[\Pi_1]=f\cdot[\Pi_0]=0$. This means that we can safely work, for instance,  with quantities involving 1-forms, which in turn imply that geometry is globally well-defined. However for D2-branes the situation is similar as D3-branes in a T-dual configuration, this is, one can not wrap D2-branes on a 2-cycle on which $N$ units of metric flux is supported unless there are $N$ D2-branes ending at it.  Hence, if we would like to define the geometry of the twisted torus in terms of  2-forms we would face a problem, since such forms are not closed. This happens for instance in the half flat manifold, which is the mirror symmetry of a Calabi-Yau manifold threaded with electric NS-NS-flux \cite{Gurrieri:2002wz, Gurrieri:2002iw}, where the K\"ahler forms are not anymore closed.\\



As in the case of NS-NS-flux, the anomalous brane supporting  metric flux can be localized in time by wrapping for instance, an E2-brane in a three-cycle in $T^6$.  In such situation, a D2-brane would disappear by encountering the E2-brane, provided there is (at least) one unit of metric flux. In the process the D2-branes transform into the flux $f\cdot F_7$, with $F_7$ being the magnetic field strength of the instantonic brane. Details of this case were studied in \cite{LoaizaBrito:2006se}.\\




\subsection{Q-twisted homology}
Incorporation of extra fluxes in a compactification scheme has notably enriched the degrees of freedom by means of which it is possible to stabilize all moduli. In particular, as was noticed in \cite{Shelton:2005cf, Shelton:2006fd, Wecht:2007wu} and later on in \cite{Aldazabal:2006up},  extra fluxes are required if one wants to keep the corresponding superpotential invariant under T-duality.  \\

In addition we want to incorporate D-branes and elucidate the conditions on which they are fully stable or unstable to transform into fluxes. Hence, in this section we study the role played by non-geometric fluxes in the stability of D-branes as well as the absence of anomalies in the world volume of the corresponding branes. D-branes in non-geometric backgrounds have been studied in \cite{Lawrence:2006ma} and in \cite{Albertsson:2008gq}, this latter  in the context of the doubled-torus formalism (see \cite{Hull:2004in, Hull:2005hk, Hull:2009sg}).\\

Let us start by briefly describing the nature of non-geometric fluxes from the perspective of T-duality.  Traditionally, non-geometric fluxes arise by performing T-duality on coordinates $x^b$ and $x^c$ on a six-dimensional torus threaded with NS-NS-flux $H_{abc}$. The resulting metric is not flat, but is not globally defined,
\begin{align}
ds^2=(dx^1)^2+\dots +\frac{1}{1+(Q^{ab}_cx^c)^2}((dx^a)^2+(dx^b)^2) + (dx^c)^2+\dots +(dx^6)^2,
\label{ngg}
\end{align}
where $Q^{bc}_a$ denotes the components of the T-dual version of $H_{abc}$ and denoted {\it non-geometric fluxes}.\\

As we said, the metric is not globally but locally defined, in the sense that local patches of the internal space are related by elements of the extended T-duality group $O(6,6;\IZ)$ \cite{Hull:2004in}. However, as opposed to the case of metric fluxes, here there is a background with non-trivial NS-NS-flux given by
$B_{ab}=\frac{Q^{ab}_cx^c}{1+(Q^{ab}_cx^c)^2}$,
where the $B$-field and the metric are mixed.\\

Before incorporating D-branes, the first thing one should notice is that the anomalous D3-brane in type IIB with NS-NS-flux T-dualizes into D1-branes supported on one-cycles on which the corresponding non-geometric flux is supported. This means that we cannot consistently define global 1-forms,  and consequently 
the metric  Eq.(\ref{ngg}) is ill-defined. 
Then, the most one can do is to define a (local) 1- form as\footnote{Which is well define up to extended T-dual maps.},
$Q_{(ab)}= Q^{ab}_c dx^c$.\\

However, as we saw in the case of NS-NS-flux, there is a way to cancel the field-theory anomaly by adding extra branes. In the present situation this translates to the fact that D3-branes extend on directions which do not support non-geometric Q-fluxes, and end at the anomalous D1-brane\footnote{By this we mean that one coordinate on D3 ends at D1.}. Since an anomalous-free configuration of D-branes involves D-branes of different dimensions, we require an extended concept of cycle which take into account all the above factors. Our goal is then to correctly define the corresponding twisted Q-homology.\\

Let us start by defining the $Q$-twisted boundary as
\begin{align}
\partial_Q=\partial+Q_{(ab)}\cap,
\end{align}
where the action of $Q$-fluxes on a $p$-chain is defined by
\begin{align}
\partial_Q\Pi_p:=\partial\Pi_p+Q_{(ab)}\cap \Pi_{p(ab)},
\end{align}
and the cap product between $Q_{(ab)}$ and the $(p+2)$-chain $\Pi_{p(ab)}$ is given by
\begin{eqnarray}
Q\cdot\Pi_p=Q_{(ab)}\cap\Pi_{p(ab)}&=&(Q^{ab}_cdx^c)\cap(\Pi_{ab}^{\mu_1\cdots\mu_{p+2}}\sigma_{\mu_1\cdots\mu_{p+2}})\nonumber\\
&=&Q^{ab}_c\tilde\Pi_{ab}^c\hat{\Pi}^{\nu_1\cdots\nu_{p+1}}\sigma_{\nu_1\cdots\nu_{p+1}}\\\nonumber
&=&N\Sigma_{p+1},
\end{eqnarray}
with $\Sigma_{p+1}$ denoting the $(p+1)$-chain $\hat{\Pi}^{\nu_1\cdots\nu_{p+1}}\sigma_{\nu_1\cdots\nu_{p+1}}$ and $\{a,b\}\in\{\mu_1\dots\mu_{p+2}\}$. Note as well that nilpotency of $\partial_Q$ is guaranteed  as far as 2$Q\cup Q=0$ (see \cite{Collinucci:2006ug}). Hence one should look at the 2-degree cohomology of local patches of $T^6$. However, the inclusion of fluxes in $T^6$ contributes positively to the tadpole. In order to cancel such positive D3-brane charge is necessary to include $O3^-$-planes. In such situation, the local patches of $T^6$ plus the orientifold planes is given by the projective space $\IRP^5$. Hence, we should look at the group $H^2(\IRP^5,\IZ)$ which turns out to be $\IZ_2$. Now, since we are working with T-dual images of NS-NS-fluxes, which components we have considered torsion-free and even-integers,  the components of the Q-fluxes are also integer numbers with no torsion\footnote{The torsion part leads to the presence of exotic orientifolds \cite{Frey:2002hf}.}, and we can safely consider that $Q\cup Q=0$.\\

Hence, as in the case of $H$-flux, we must take into account  a net of cycles and chains in order to define a $Q$-closed cycle.  According to the above equations we see that
\begin{align}
\partial_Q[\Pi]=\partial_Q([\Pi_p]-N\Theta_{p+2})=0,
\end{align}
where $\partial\Theta_{p+2}=[\Sigma_{p+1}]$, and
\begin{align}
\partial[\Pi_p]=0\nonumber\\
Q_{(ab)}\cap \Theta_{p+2(ab)}=0.
\end{align}
The generalized cycle $[\Pi]$ is globally well defined and allows to wrap, for instance, D1-branes on coordinates where the non-geometric flux is defined, as far as there are extra D3-branes extending other coordinates. Let us illustrate such configurations by a concrete example.\\

\noindent
{\bf Example.} Take IIB string theory compactified on a factorizable $T^6$ threaded with a non-geometric flux $Q_{(12)}=Q^{12}_3dx^3$. Consider as well the 1-cycle given by
\begin{align}
[\Pi_1]=\Pi^3\sigma_3,
\end{align}
such that $\partial[\Pi_1]=0$. We can see that this is not a $Q$-cycle since
\begin{eqnarray}
\partial_Q[\Pi_1]&=&Q_{(12)}\cap[\Pi_{1(12)}]\nonumber\\
&=&Q^{12}_3dx^3\cap \Pi_{12}^{312}\sigma_{312}\nonumber\\
&=&Q^{12}_3\Pi_{12}^3\tilde{\Pi}^{12}\Sigma_{12}\nonumber\\
&=&N[\Pi_2]=N(\Pi^{12}\sigma_{12}).
\end{eqnarray}
On the other hand, let us denote a 3-cycle $\Sigma_3$ as $\Sigma^{\delta\mu\nu}\sigma_{\delta\mu\nu}$ with coordinate\footnote{The reader must not be confused about the index 3 in $[\Sigma_3]$ which denotes the dimension of the chain, and the coordinate $x^3$ which does not belong to $x^\delta,x^\mu, x^\nu$. Also notice that $[\Pi_{1(12)}]$ denotes the extension of $[\Pi_1]$ to coordinates (12).} $3\neq\{ \delta\mu\nu\}$. 
Hence, if one considers $N$ extra D3-branes wrapping $\Sigma_3$ such that $\partial\Sigma_3=[\Pi_2]$ we get that  
\begin{align}
\partial_Q[\Pi]=\partial_Q([\Pi_1]-N\Sigma_3)= 0,
\end{align}
for $\Sigma_3=\Sigma^{12z}\sigma_{12z}$ where $z$ denotes an extended spatial coordinate, i.e., $\Sigma_3$ cannot be an internal 3-chain.\\

We see then that it is possible to wrap D1-branes on coordinate $x^3$ in the presence of $N$ units of Q-flux $Q_{(12)}$ as far as there are $N$ D3-branes wrapping an internal 2-cycle and one extended direction. Although we shall not study the R-space, it is interesting to point out that according to our results, one can define a point in such space as far as there are D4-branes wrapping internal three-cycles and one extended coordinate.\\

\subsubsection{Instantonic branes in Q-space}
Likewise to H-space and f-space, we expect that non-geometric fluxes drive some branes to decay into fluxes. In this section we study the details of such configuration by considering a concrete example.

Take for instance a E1-brane localized in time,  wrapping a 2-cycle $[\Pi_2]=\Pi^{3m}\sigma_{3m}$ in a non-geometric background given by the local Q-flux  $Q_{(12)}=Q^{12}_3dx^3$. The $Q$-cycle turns to be 
\begin{align}
[\Pi]=[\Pi_2]-N\Sigma_4,
\end{align}
with $\Sigma_4=(\Sigma^{12\mu}\sigma_{12\mu})\times (-\infty, t_0]$, where $t_0$ is the time at which the instantonic E1-brane appears. Hence, $N$ D3-branes become unstable and transform into flux. The remnant flux is given by the non-geometric fluxes coupled with the magnetic field strength of $E1$, i.e., the RR flux $F_7$.
After disappearing the $N$ D3-branes transform into (see \cite{Shelton:2005cf} for the action of $Q$ on $p$-forms)
\begin{align}
Q_{(12)}\cdot F_7:=Q^{12}_3F_{12xywzk}dx^3\wedge\cdots\wedge dx^k=dF_5,
\end{align}
and such a flux is closed under $d_Q$, this is
\begin{align}
dF_5-Q\cdot F_7=0,
\end{align}
which tells us that D3-brane charge is carried by the composite flux $Q\cdot F_7$. \\

Notice that, as in  previous cases, the tadpole condition is not modified since the disappearing branes become fluxes. Specifically, in the presence of non-geometric fluxes the tadpole condition reads \cite{Aldazabal:2006up}, 
\begin{align}
{\cal N}_{D3} + Q\cdot F_7=0.
\end{align}

In the following section, we shall consider a more 'realistic' model, involving D3-branes sitting at a point in the internal six-dimensional non-geometric torus.

\section{Effects of brane-flux transition on SUSY vacuum}
As we have seen, the presence of fluxes  (3-forms, metric and non-geometric ones),  drive some D-branes to decay into a configuration of fluxes which consists in 
a coupling
between (the magnetic field strength) RR flux of the instantonic brane and the above mentioned fluxes. In such situation, it is the total number of (corresponding) D-branes plus the number of flux which is invariant under the appearance of instantonic E-branes. This is precisely what counts on the tadpole condition on flux compactifications.  In this section we shall  see that,  although tadpole condition is invariant under brane-flux transitions, quantities depending on RR fluxes are not. One example is the Gukov-Vafa-Witten superpotential \cite{Gukov:1999ya} and its generalization to non-gometric compactifications. The transition between branes into fluxes changes the superpotential and hence, changes the SUSY vacuum solutions derived from it.  This kind of transitions between flux vacuum was first studied in \cite{Kachru:2002he} (see also \cite{deAlwis:2006cb} in the context of supergravity). In this section we shall recover such results from the perspective of brane-flux transitions mediated by instantonic branes and we shall describe the generalizations to the case of non-geometric flux backgrounds. 

\subsection{Vacuum transitions in 3-form flux tori compactifications}
Consider type IIB string theory compactified on a six-dimensional factorizable torus $T^6=\otimes_{i=1}^3T_i^2$ in an $O3^-$-plane background threaded with NS-NS and RR 3-form fluxes.
Adopting  the notation in \cite{Aldazabal:2006up}, the basis of  closed 3-forms in $T^6$,  classified by $\text{H}^3(T^6;\IZ)=\IZ^8$,  is writen as
 \begin{align}
 \alpha_0=dx^1\wedge dx^2\wedge dx^3, \quad\quad \quad \beta_0=dx^4\wedge dx^5\wedge dx^6,\nonumber\\
 \alpha_1=dx^1\wedge dx^5\wedge dx^6, \quad\quad\quad  \beta_1=dx^4\wedge dx^2\wedge dx^3,\nonumber\\
 \alpha_2=dx^4\wedge dx^2\wedge dx^6, \quad\quad\quad \beta_2=dx^1\wedge dx^5 \wedge dx^3,\nonumber\\
 \alpha_3=dx^4\wedge dx^5\wedge dx^3, \quad\quad\quad \beta_3=dx^1\wedge dx^2 \wedge dx^6,
 \end{align}
such that $\int_{T^6}\alpha_I\wedge\beta_J=\delta_{IJ}$ and where the torus $T_i^2$ is spanned by coordinates $(x^i, x^{i+3})$. The corresponding cycles are denoted as
\begin{align}
{\cal PD}(\alpha_0)=\sigma_{456}, \quad\quad\quad {\cal PD}(\beta_0)=\sigma_{123},\nonumber\\
{\cal PD}(\alpha_1)=\sigma_{423}, \quad\quad\quad {\cal PD}(\beta_1)=\sigma_{156},\nonumber\\
{\cal PD}(\alpha_2)=\sigma_{153}, \quad\quad\quad {\cal PD}(\beta_2)=\sigma_{426},\nonumber\\
{\cal PD}(\alpha_3)=\sigma_{126}, \quad\quad\quad {\cal PD}(\beta_3)=\sigma_{453}.
\end{align}
For simplicity we are going to consider identical two-dimensional torus, such that the complex structure $\tau$ is the same for all three tori. In this context, the holomorphic 3-form $\Omega$ is given by
\begin{eqnarray}
\Omega(\tau)&=&\wedge_{i=1}^3(dx^i+iU dx^{i+3}),\nonumber\\
&=& \alpha_0+iU\sum_i \beta_i -U^2\sum_i\alpha_i-iU^3\beta_0,
\label{omega}
\end{eqnarray}
and the 3-form fluxes are given in their most general form by\footnote{Along this paper, we take $(2\pi^2)\alpha'=1$.}
\begin{align}
F_3= -m\alpha_0-e_0\beta_0+\sum_{i=1}^3(e_i\alpha_i-q_i\beta_i)\nonumber\\
H_3=h_0\beta_0+\bar{h}_0\alpha_0-\sum_{i=1}^3(a_i\alpha_i+\bar{a}_i\beta_i).
\label{HF1}
\end{align}
In such situation, as mentioned, 3-form fluxes induce a $D3$-brane charge in the compact manifold, measured by the flux number
\begin{align}
{\cal N}_{\text{flux}}=\int H_3\wedge F_3= h_0m-e_0\bar{h}_0+\sum_i(a_iq_i+e_i\bar{a}_i).
\end{align}
In order to avoid orientifold subtleties \cite{Frey:2002hf}, we shall concentrate only on even fluxes. The orientifold $O3^-$-plane cancels the $D3$-brane charge generated by the fluxes and physical $D3$-branes, such that the total contribution denoted  $[{\cal N}_{\text{$D3$/flux}}]$ equals the $D3$-brane charge of the orientifold, i.e. $
[{\cal N}_{\text{$D3$/flux}}]= 16$,
where
\begin{align}
[{\cal N}_{\text{$D3$/flux}}]={\cal N}_{D3}+\frac{1}{2}(h_0m-e_0\bar{h}_0+\sum_i(a_iq_i+e_i\bar{a}_i)).
\end{align}
On the other hand, as it is well known \cite{Gukov:1999ya}, the 3-form fluxes give rise to a superpotential ${\cal W}_H=\int_{T^6}(F_3-SH_3)\wedge \Omega(\tau)$, where the complex dilaton is $S=C_0+ie^{-\phi}$. Let us simplify the conditions and  take the isotropic case in which $e_i=e$, $q_i=q$, $a_i=a$ and $h_i=h$ for all $i$, thus the no-scale superpotential ${\cal W}_H(\tau, S)$ in terms of the symplectic basis $(\alpha_I,\beta_I)$ reads
\begin{align}
{\cal W}_H=(e_0+Sh_0)+ 3i(e+Sa)U-3(q-S\bar{a})U^2 +  i (\bar{h}_0S+m)U^3,
\label{wh}
\end{align}
which can be written in terms of two polynomial as
\begin{align}
{\cal W}_H= P_1(\rho)-SP_2(\rho),
\end{align}
with $U=-i\rho$ and
\begin{align}
P_1(\rho)= e_0+3e\rho+3q\rho^2-m\rho^3,\nonumber\\
P_2(\rho)= -h_0-3a\rho +3\bar{a}\rho^2+\bar{h}_0\rho^3.
\end{align}
At this point, we are interested in  supersymmetric Minkowsky vacuum solutions\footnote{Being $\W_H$ a no-scale superpotential, any supersymmetric solution would be Minkowski.}, for which the corresponding equations read $\partial_S\W_H=\partial_\rho\W_H=\W_H=0$. A solution for these equations leads to the existence of a common root for $P_1(\rho)$ and $P_2(\rho)$ denoted $P(\rho)$ from which a complex value for $\rho$ is computed, i.e., 
\begin{align}
P_i(\rho)=P(\rho)(f_i\rho + g_i),
\end{align}
with $f_i$ and $g_i$ integers ($i=1,2$) and $P(\rho)=(\rho-\rho_0)(\rho-\rho_0^\ast)$. Hence the complex structure is stabilized at the value $U=-i\rho_0$ such that $Re(U)>0$. Notice that this solution comes from the pair of equations $\W_H=\partial_S\W_H=0$ and important enough, that even though there are solutions for which $P_1(\rho_1)=P_2(\rho_2)=0$, if  there is no common root, i.e. $\rho_1\neq\rho_2$ the solution is inconsistent and it does not represent a supersymmetric solution.\\

From $\partial_\rho\W_H=0$ one gets the value of the complex dilaton (with $Re(S)>0$ as well), which fulfills the equation
\begin{align}
P_1'(\rho)-SP_2'(\rho)=0,
\end{align}
with $P_i'=\partial_\rho P_i$. Notice that the pair of integers $(f_1, g_1)$ and $(f_2,g_2)$ depend only on the number of  units of RR and NS-NS fluxes respectively. \\

We have now all the ingredients to study the effects of the brane-flux transformation on the supersymmetric vacuum solutions described in  previous section . Let us start by studying how brane-flux transformation works on a very simple case in which the NS-NS flux is given by $H_3=-a\alpha_2$  (i.e., $H_{426}=-a)$, $F_3=-q\beta_2$ and  with $N$ D3-branes sitting at a point in $T^6$. Up to the appearance of an instantonic E5-brane wrapping the extended spatial coordinates $x,y,z$ and the internal 3-cycle $\sigma_{426}$, some D3-branes would be unstable to decay into fluxes. 
Hence, $-a$ D3-branes out of the originally $N$ transform into one unit of RR flux $F_{153}$ coupled with the already present NS-NS flux $H_3$, i.e, the amount of RR flux $F_{153}$ increases by one unit. However,  we must take into account the presence of the  orientifold 3-plane. In such case, since the instantonic E5-brane makes unstable the D3-branes and their images under the $O3^-$-plane, the correct receipt reads that $-a/2$ D3-brane become $-a$ units of flux $F_3\wedge H_3=-adx^1\wedge\cdots\wedge dx^6$ and the final RR flux is $F_3=(-q+1)\beta_2$. It follows that $[\N_{\text{$D3$/flux}}]$  keeps invariant since after the transition, the tadpole condition reads
\begin{align}
(N+\frac{a}{2})+\frac{a}{2}(q-1)=[{\cal N}_{\text{$D3$/flux}}].
\end{align}
Notice that the number $a$ must be an even integer in order to avoid fractional D-branes. This is translated after the transformation from branes into fluxes, to the well known fact that we are considering even fluxes in order to avoid subtleties with exotic orientifold planes \cite{Frey:2002hf}.\\

\noindent
{\bf Example.} The effects on supersymmetric solutions are easy to obtain, and we shall study them in a very well known example of SUSY solution in a torus flux compactification described in \cite{Kachru:2002he}.
Take the (non S-dual) flux background,
\begin{align}
F_3= 2\alpha_0+2\beta_0,\nonumber\\
H_3=-4\beta_0+2\alpha_0+2\sum_i(\alpha_i-\beta_i),
\label{HF}
\end{align}
which produces the superpotential $\W_H=P_1(\rho)-SP_2(\rho)$ with
\begin{align}
P_1(\rho)= 2(\rho^3-1)=2(\rho-1)(\rho^2+\rho+1),\nonumber\\
P_2(\rho)= 2(\rho^3+3\rho^2+3\rho +2)=2(\rho+2)(\rho^2+\rho+1).
\end{align}
The common root $P(\rho)=\rho^2+\rho+1$ gives the value for complex structure $\rho_0=e^{\frac{2\pi}{3}i}$. The value of the dilaton is in this case $S=\rho_0$.  Notice that the complex structure vev is computed from the polynomial $P(\rho)$ which does not depend on the amount of RR fluxes. From the tadpole condition, one realizes that there must be 10 D3-branes sitting at a point in $T^6$ and  ${\cal N}_{\text{flux}} = 12$.\\

\begin{figure}[t]
\begin{center}
\centering \epsfysize=8cm \leavevmode
\epsfbox{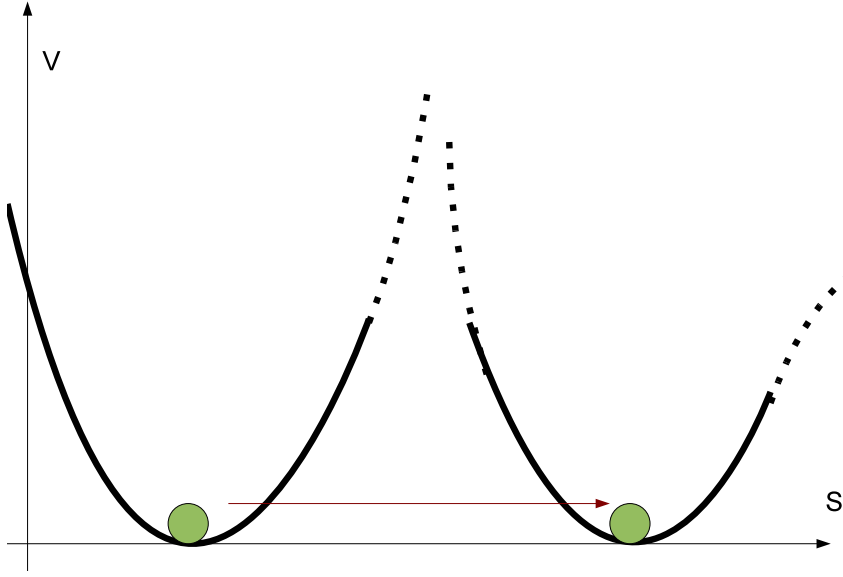}
\end{center}
\caption[]{\small \it Two different supersymmetric vaccua connected by a brane-flux transition, in which $P_1\rightarrow \kappa P_1$ and $S\rightarrow \kappa S$.} 
\label{kachru}
\end{figure}

Now, upon the appearance of instantonic E5-branes wrapping internal cycles some D3-branes would transform into fluxes.  Actually for two, separated in time,  E5-branes wrapping the cycle $\sigma_{456}$, 4 D3-branes would disappear leaving as a remnant 2 extra units of RR flux $F_{ 123}$ while for two E5-brane wrapped on $\sigma_{123}$ 2 D3-branes would transform into flux increasing the units of RR flux on the 456 component by 2 units. Afterwards, correctly counting the orientifold D3-brane images as well, the fluxes become
\begin{align}
F_3=4\alpha_0+4\beta_0,
\end{align}
with $H_3$ the same as before. Notice that $\Nf$ is invariant, but the superpotential $\W_H$ has changed. The modification is reflected in $P_1(\rho)$ which now reads
\begin{align}
P_1(\rho)= 4(\rho^3-1).
\end{align}
It follows that the complex structure keeps its value at $\rho_0=e^{\frac{2\pi}{3}i}$ whilst the dilaton moves its value at $S=2\rho_0$. At this point we have connected 2 different vacuum solutions (with different values for the dilaton) through out the appearance of an instantonic brane. The amount of flux have increased while the number of D3-branes has reduced to four. 
For this case, we can perform more transitions to reach a pure flux configuration. Along the way, the dilaton increases its value by some integers.
So, it seems that by rescaling the units of RR flux, we can connect different supersymmetric vacuum solutions (Figure \ref{kachru}) . This fact was as well studied in \cite{Kachru:2002he} where the authors found a general set of connection in terms of $GL(6,\IZ)\times GL(2,\IZ)$ transformations. Here, we have recovered such results in the context of brane/flux transitions mediated by an instantonic E5-brane.\\

\begin{figure}[t]
\begin{center}
\centering \epsfysize=10cm \leavevmode
\epsfbox{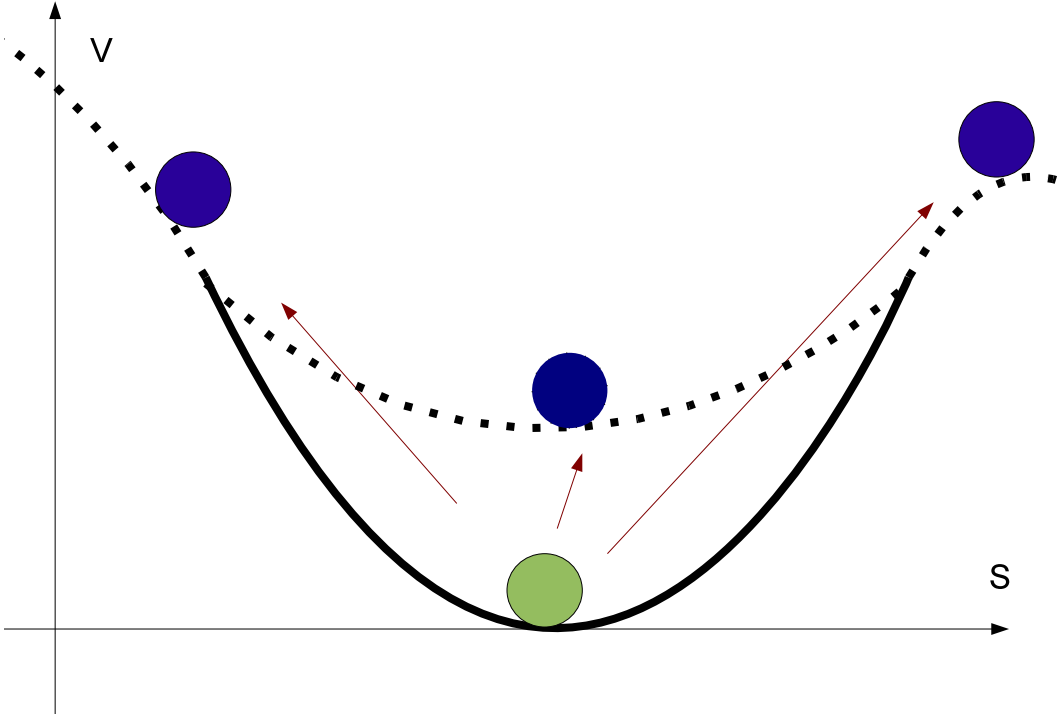}
\end{center}
\caption[]{\small \it Under a brane-flux transition,  a supersymmetric vacuum ( solid line) is connected to  a non-supersymmetric flux configuration, in which the scalar potential (dotted line) could have a minimum.} 
\label{nonsusy}
\end{figure}

However, as we can see from the original set of fluxes (\ref{HF}), there are six more three-cycles on which instantonic E5-branes could be  wrapped supporting NS-NS-flux. For E5-branes wrapping cycles ${\cal PD}(\alpha_i)$ and ${\cal PD}(\beta_i)$, one single D3-brane transforms into flux for each cycle. Hence, by considering a process for each possible instantonic brane, it may happen that 6 D3-branes become flux and that the set of RR fluxes in (\ref{HF}) changes into
\begin{align}
F_3=2\alpha_0+2\beta_0+\sum_i(\alpha_i+\beta_i),
\end{align}
i.e., $-q=0 \rightarrow -q=1$ and $e=0\rightarrow e=1$
from which the polynomial $P_1(\rho)$ reads now
\begin{align}
P_1(\rho)  \rightarrow P_1(\rho)=  2\rho^3-3\rho^2+3\rho-2.
\end{align}
The root for the polynomial $P_1(\rho)$ is not the same at which the complex structure stabilizes in the starting configuration. Actually $P_1(\rho_\ast)=0$ for $\rho_\ast= \frac{1-i\sqrt{15}}{4}$. Hence, at $\rho_0=e^{\frac{2\pi}{3}i}$, $P_1(\rho_0)\neq 0$, implying that $\W_H(\rho_0)\neq 0$ and thus, by the appearance  of instantonic branes, supersymmetry is broken.  The breakdown is mediated through all moduli, since now $D_S\W, D_T\W$ and $D_\rho\W$ are all different from zero. Notice however that for the moduli to be stabilized at the point in moduli space $\rho=\rho_*$, the scalar potential should be at a minimum above zero. A more detailed analysis is required to prove this and we left it to further work. It is sufficient here to realize that some transitions are just in general moving the moduli from values at which they are stabilized in a supersymmetric configuration into ones in which supersymmetry is broken and (perhaps) the scalar potential does not posses a minimum (see Figure \ref{nonsusy}).\\


Having studied a concrete example, let us discuss a general case. 
Take the set of fluxes (\ref{HF1}). For each integer in the expansion for the NS-NS flux (there are 8 different fluxes reduced to 4 in the isotropic case), the appearance of instantonic branes wrapping suitable cycles would increase the number of corresponding RR fluxes by one unit as,
\begin{eqnarray}
-h_0&:& -m \rightarrow -m+1\nonumber\\
\bar{h}_0&:& -e_0\rightarrow -e_0+1\nonumber\\
-a_i&:&  -q_i\rightarrow  -q_i+1\nonumber\\
-\bar{a}_i&:& ~~ e_i\rightarrow ~~ e_i+1.
\end{eqnarray}
Tadpole condition $[\Nf]$ for this general case also keeps invariant under the brane-flux transformation, since
\begin{eqnarray}
[\Nf]&\rightarrow& (N-\frac{1}{2}(h_0+\sum_i(a_i+\bar{a}_i)+\bar{h}_0))\nonumber\\
&&+\frac{1}{2}\left(h_0(m-1)-\bar{h}_0(e_0-1)+\sum_i[a_i(1-q_i)+\bar{a}_i(e_i+1)]\right) \nonumber\\
&=&[{\cal N}_{D3\text{/flux}}],
\end{eqnarray}
but the superpotential $\W_H$ is not. A chain of  instantonic branes wrapping all cycles on which NS-NS-flux is supported would generate changes in the superpotential. Let us denote the number of instantonic branes by $\kappa_i$ where $i$  stands for a specific cycle. Then, for instance, for the isotropic case the superpotential written in terms of RR fluxes $(m, e_0, e, q)$ would transform as
\begin{eqnarray}
&&{\W}_H(\dots,-m, -q, -e_0, e,\dots)\leftrightarrow\nonumber \\
&&{\W}_H(\dots, -m+\kappa_1, -q+(\kappa_2+\kappa_3+\kappa_4), -e_0+\kappa_5, e+(\kappa_6+\kappa_7+\kappa_8),\dots).\nonumber
\end{eqnarray}
Observe that since brane-flux transition acts only on RR fields, it would then modify the polynomial $P_1(\rho)$ keeping $P_2(\rho)$ identical.
Before the transition, the complex structure $\rho$ is stabilized at the root $\rho_0$ of $P(\rho)$, and the dilaton $S$ fulfills the equation $P_1'(\rho_0)-SP_2'(\rho_0)=0$. After the transition we have then two general cases:
\begin{enumerate}
\item
$P_1(\rho)=P(\rho)Q_1(\rho)\rightarrow \hat{P}_1(\rho)=P(\rho)\hat{Q}_1(\tilde{f}_1\rho+\tilde{g}_1)$, where $\tilde{f}$ and $\tilde{g}$ are functions of the modified RR fluxes. The complex structure is invariant (there is still a common root between $P_2$ and $\hat{P}_1$), but the value at which the dilaton is stabilized suffers a change. The prototypical example for this case takes a transformation on $P_1(\rho)$ to $\kappa P_1(\rho)$ from which one gets that the complex structure is stabilized at the same value $\rho_0$.  However, from the F-flat equation $\partial_\rho\W_H=0$ we get that the dilaton $S$ must fullfill the equation
$\kappa P_1'(\rho)- S P_2'(\rho)=0$,
which is true for $S=\kappa \rho_0$.
\item
The transition changes the polynomial such that there is a no common factor $P(\rho)$. The new polynomial  reads $\hat{P}_1(\rho)=\hat{P}(\rho)\hat{Q}_1(\rho)$ with $\hat{P}(\rho) =0$ at $\rho=\rho_*(\neq \rho_0)$. Hence, at $\rho=\rho_\ast$  and at $\rho=\rho_0$, $\W_H\neq 0$ from which we see that supersymmetry is broken through all moduli. At both points, the scalar potential is positive (since the model is no-scale) but an extended and more detailed analysis is required to elucidate if the scalar potential is minimum. In the case in which the scalar potential is not minimum at those points, we can say that brane-flux transitions connect the supersymmetric configuration into some unstable excitations around the basis point.  See Figure \ref{nonsusy}.
\end{enumerate}

The most common and most studied case is the one described in point 1. This type of transition has been studied in the past in an specific geometric configuration as the singular points in a conifold, where a geometric transition connects two different configurations where branes and fluxes interchange their roles (see for instance section 4.2 in  \cite{Blumenhagen:2009qh}). At the level of supergravity, the transition were also studied in \cite{deAlwis:2006cb}. \\

Mathematically speaking, different configurations belong to the same class of configuration as classified by K-theory\footnote{The transition between branes and fluxes, as seen in section 2, is the physical realization of the Atiyah-Heirzebruch Spectral Sequence, which is an algorithm connecting cohomology classes with K-theory ones \cite{Maldacena:2001xj}.} and labeled by the number of flux-brane contribution to the $D3$-brane charge, i.e., by $[\Nf$]. However, as we have seen, there are more possible fluxes to consider in a string compactification. Our next step is to study the effects of such transitions on S-dual versions and on non-geometric compactifications.

\subsection{S-dual invariance and extra fluxes}
As it is well known, type IIB string theory is invariant under $SL(2,\IZ)$ transformations. In particular the flux configuration $H_3\wedge F_3$ is dual, leading us to the expectation that a transition between branes and fluxes also obey this symmetry \cite{Evslin:2003hd}. Although not formally proved, an S-dual version of such transformation has been already used in different scenarios (\cite{LoaizaBrito:2003gz, Evslin:2004vs, Bouwknegt:2005ky}, see also (\cite{Villadoro:2007yq, Villadoro:2007tb}).\\

Roughly speaking, the proposal establishes that RR 3-form fluxes also drive some D3-branes to be unstable via the nucleation of an instantonic NS5-brane\footnote{Nucleation of NS5-branes has been used, for instance, in \cite{Kachru:2002gs}.}. Essentially, the S-dual version of twisted Bianchi identities read
\begin{align}
d_F(F_5-H_3)=0,
\end{align}
with $d_F=d-F_3\wedge$ acting on poli-forms as well, which is the S-dual version of $d_H$. It is then straightforward to define the S-dual version of the twisted homology cycle as,
\begin{align}
\partial_F= \partial - F_3\cap.
\end{align}
Thus, an F-cycle is constructed by a NS5-brane which supports non-trivial units of RR flux $F_3$ and $D3$-branes ending at it. At the level of field theory, this gives the required monopole for the induced gauge fields on the NS5-brane worldvolume.
By fixing the NS5-brane in time, we get that some $D3$-branes would be unstable to decay into fluxes when RR fluxes are present. We require, as in the original configuration,  that the pullback of the RR 3-form flux $F_3$ does not vanish on the NS5-brane worldvolume.\\

Thus, there are actually two ways by which D3-branes transform into fluxes and even more, there is a transition by which the units of fluxes, both RR and NS-NS 3-forms, increase (or decrease). This has some very important consequences in the construction of a Minkowski vacuum in tori compactifications, since, as we have seen, if the amount of NS-NS flux is altered, the corresponding superpotential $\W_{(H,F)}$ would change via the polynomial $P_2(\rho)$. \\ 

\subsection{S-dual vacuum transitions in 3-form fluxes tori compactifications}
As in the case of transitions between branes and fluxes via instantonic E5-branes, an instantonic NS5-brane supporting $N$ units of RR flux $F_3$ would drive $N$ D3-branes to decay into fluxes, which consists in the wedge product of the already present RR flux $F_3$ and one unit of NS-NS flux representing the magnetic field strength of the disappearing NS5-brane. From the general set of fluxes (\ref{HF1}) we see that for each non-zero unit of RR flux, there will be a change in the corresponding units of  the NS-NS flux, according to:
\begin{align}
-m: h_0\rightarrow h_0+1\nonumber\\
e_0: \bar{h}_0\rightarrow \bar{h}_0+1\nonumber\\
e_i:-\bar{a}_i\rightarrow -\bar{a}_i+1\nonumber\\
q_i: -a_i\rightarrow -a_i+1.
\label{nsnschanges}
\end{align}
Hence, upon the nucleation of NS5-branes there will be changes in the set of fluxes, implying a transition on the vacuum solutions via the flux superpotential.\\

\noindent
{\bf Example.}
In order to clarify ideas, let us consider the example studied in section 3.1 allowing in this case the nucleation of instantonic NS5-branes. The original set of fluxes (\ref{HF}) involves 2 units of RR flux on the basis $\alpha_0$ and $\beta_0$ respectively which can source the instability of certain D3-branes. First, consider the appearance of 2 instantonic E5-branes along internal cycles ${\cal PD}(\alpha_1)=\sigma_{423}$ and ${\cal PD}(\beta_1)=\sigma_{156}$. As we have seen, this would drive 4 D3-branes to decay into fluxes. Magnetic fields related to the instantonic branes increase the units of RR 3-form flux to
\begin{align}
F_3 \rightarrow 2\alpha_0+2\beta_0+\beta_1+\alpha_1.
\label{fns}
\end{align}
Now, let us say that 3 NS5-branes nucleate wrapping the internal cycles $\sigma_{243}$ (twice) and  $\sigma_{156}$. In this case the remnants affect the units of NS-NS flux which according to  (\ref{nsnschanges}) transform into\footnote{Strictly speaking, although the fluxes $F_3$ and $H_3$ are even integer fluxes, if we analyze the amount of flux on each cycle, RR fluxes in (\ref{fns}) are odd in $\beta_1$ and $\alpha_1$. Therefore, fractional branes decay upon the appearance of intantonic branes wrapping $\sigma_{423}$ and $\sigma_{156}$. In order to posses an even number of flux on each cycle, we must consider an equivalent flux configuration as $F_3=2(\alpha_0+\beta_0+\alpha_1+\beta_1)-\alpha_2-\beta_2$, and we are not considering a transition involving $\beta_2$ and $\alpha_2$.} 
\begin{align}
H_3\rightarrow -4\beta_0+2\alpha_0-\beta_1 +2(\alpha_2+\alpha_3-\beta_2-\beta_3).
\label{hns}
\end{align}
Fluxes (\ref{fns}) and (\ref{hns}) fix the new superpotential $\W_{(F,H)}$ through the polynomials
\begin{align}
P_1(\rho)= 2\rho^3-\rho^2+\rho-2,\nonumber\\
P_2(\rho)= 2\rho^3+5\rho^2+4\rho+4,
\end{align}
which have a common factor
\begin{align}
Q(\rho)= 2\rho^2+\rho+2.
\end{align}
This fixes the complex structure at 
\begin{align}
\rho_\ast=-\frac{1}{4}(1+i\sqrt{15}),
\end{align}
and the dilaton at
\begin{align}
S(\rho_\ast)=-\frac{1}{16}(5+3i\sqrt{15}).
\end{align}


We observe that the transition changes the polynomial $P_1(\rho)$ and $P_2(\rho)$  such that there is still a common root since they share a common factor.  However, this common factor $Q(\rho)$ has also changed with respect to $P(\rho)$ gathered from the original set of fluxes. The new polynomials are  $P_1(\rho)= Q(\rho)(\tilde{f}_1\rho +\tilde{g}_1)$ and $P_2(\rho)=Q(\rho)(\tilde{f}_2\rho + \tilde{g}_2)$ with $Q(\rho_\ast)=0$ and $\rho_\ast\neq\rho_0$. In this case, the transition connects two different supersymmetric solutions which differ not only by a rescaling of fluxes. Therefore, the nucleation of instantonic NS5-branes 
allows us to connect many different vacuum solutions which otherwise would be disconnected. The S-dual version of the brane-flux transition must  notably reduce the size of vacuum solutions.\\


\section{Effects of non-geometric  brane-flux transitions on SUSY vacuum}
We have seen in section 2  that the presence of non-geometric fluxes contribute to the tadpole in string IIB simple orientifold compactifications. This means that D-brane charge is also carried by non-geometric fluxes coupled with appropriate RR ones.  In this section we shall study the effects of non-geometric flux compactifications in transitions between Minkowski vacuum. As in previous case, we shall concentrate on factorizable tori compactifications. However, since a Minkowski solution should fulfill many restrictions, finding a suitable example is quite complicated. We shall limit ourselves to describe some specific examples instead of offering a complete general picture.\\

\subsection{Vacuum transition in non-geometric tori compactifications}
Consider as before type IIB string theory on a simple orientifold threaded with non-geometric fluxes $Q$.  We shall closely follow notation and conventions in \cite{Aldazabal:2006up}. Hence there are 24 non-geometric fluxes which survive the orientifold projection and which are related by a chain of T-dualities to fluxes in type IIA with $O6$-planes and type IIB in an  $O9$-plane. They are arranged in the two matrices
\begin{equation}
\mathbf{b}=\left(
\begin{array}{ccc}
-Q^{23}_1 & Q^{34}_5 & Q^{42}_6\\
Q^{53}_4 & -Q^{31}_2 & Q^{15}_6\\
Q^{26}_4 & Q^{61}_5 & -Q^{12}_3
\end{array} \right)
\qquad 
\mathbf{\bar{b}}=\left(
\begin{array}{ccc}
-Q^{56}_4 & Q^{61}_2 & Q^{15}_3\\
Q^{26}_1 & -Q^{64}_5 & Q^{42}_3\\
Q^{53}_1 & Q^{34}_2 & -Q^{45}_6
\end{array} \right),
\label{bb}
\end{equation}
and the two vectors
\begin{align}
\mathbf{-\overrightarrow{\mathbf{h}}}=\left( Q^{23}_4, Q^{31}_5, Q^{12}_6\right) \qquad -\mathbf{\overrightarrow{\mathbf{\bar{h}}}}=\left(Q^{56}_1, Q^{64}_2, Q^{45}_3\right).
\label{hh}
\end{align}
We shall study the isotropic case $\mathbf{b}_{ij}=b$ for $i\neq j$ and $\mathbf{b}_{ii}=\beta$ and similar expressions for $\mathbf{\bar{b}}$ and $\mathbf{\bar{h}}$. \\

Now, since we are interested in the induction of RR 3-form fluxes by a brane-flux transition, we shall concentrate on couplings between $F_3$ and non-geometric fluxes $Q$. The action of $Q$-fluxes on a 3-form $F_3=\frac{1}{3!}F_{abc}dx^a\wedge dx^b\wedge dx^c$ is \cite{Wecht:2007wu},
\begin{align*}
Q\cdot F_3= Q^{mn}_pF_{mna}dx^p\wedge dx^a=dF_1,
\end{align*}
from which one sees that $Q$-fluxes couple to $F_3$ and carry D7-brane charge. The corresponding $Q$-cycle, is computed according to the rules described in section 2 (see the example below). The $F_3$ RR flux is induced by an instantonic $E5$-brane wrapped on $[\Sigma_{xyz}]$ (the non-compact 3d-space) times an internal 3-cycle $[\Pi_3]$, on which $N$-units of non-geometric flux are supported. In addition take an 8-cycle $[\Pi_8]$ (extended on $X_4$ and an internal 4-cycle $[\Pi_4]$)
on which the worldvolume of $N$ D7-branes are wrapped\footnote{The characteristics of the above cycles will be elucidated in the next example.}. The poli-cycle 
\begin{align}
[\Pi]=[\Sigma_{xyz}]\times [\Pi_3]-N[\Pi_8],
\end{align}
satisfies $\partial_Q[\Pi]= 0$ implying that $N$ D7-branes can be unstable to decay into the flux $Q\cdot F_3$. Since the total amount of brane-flux is constant, we get that in a 
compact manifold
\begin{align}
\mathscr{N}_{D7}+Q\cdot F_3=0,
\end{align}
which is actually the tadpole condition. However since there are 3 different ways to wrap a D7-brane we shall follow the notation \cite{Aldazabal:2006up}  and use $D7_i$ to denote the D7-brane transversal to $T^2_i$. Taking the RR flux  in (\ref{HF1}), it is straightforward to write the tadpole condition for the isotropic case we are considering, as
\begin{align}
[\mathscr{N}_{D7_i/\text{flux}}]=\mathscr{N}_{D7_i}-\frac{1}{2}\left[(mh_i-e_0\bar{h}_i)-q(\beta+2b)-e(\bar{\beta}+2\bar{b})\right]=0.
\end{align}
Due to the transition between D7-branes and Q-fluxes, it is clear that (after transition) there will be extra units of RR flux $F_3$. However, as in the previous case, the tadpole condition remains invariant and $[\mathscr{N}_{D7/\text{flux}}]$ always vanishes. In order to elucidate this in detail, it is necessary to find the transition for each $Q$-flux out from the 24 fluxes we have in (\ref{bb}) and (\ref{hh}). Notice that in this case there are different fluxes driving  the same $D7_i$-brane to decay into fluxes. \\

\noindent
{\bf Example.}  Consider an instantonic $E5$-brane wrapped on the six-dimensional cycle $[\Pi_6]=[\Pi_3]\times\Sigma_{xyz}$, where $\Sigma_{xyz}$ denotes the 3-dimensional uncompactified Minkowski space.  To be more specific, let us take the internal cycle as
\begin{align}
[\Pi_3]= \Pi^{123}\sigma_{123}.
\end{align}

We want to see whether non-geometric fluxes in (\ref{bb}) and (\ref{hh}) act on this internal cycle.  If so,  $Q\cdot [\Pi_3]$ must be different from zero. To show this, take the general local 1-form $Q_{(ab)}$ and act it on $[\Pi_3]$. This is 
\begin{align}
Q\cdot [\Pi_3]=Q_{(ab)}\cap [\Pi_{3(ab)}]= (Q^{ab}_cdx^c)\cap \left(\Pi^{123ab}_{ab}\sigma_{123ab}\right),
\end{align}
where $\{a,b\}\neq \{1,2,3\}$ but $c\in\{1,2,3\}$. Hence, according to the fluxes in (\ref{bb}) and (\ref{hh}), we get
\begin{align}
Q\cdot[\Pi_3]= (Q^{56}_1dx^1)\cap (\Pi^{12356}_{56}\sigma_{12356}) + (Q^{64}_2dx^2)\cap(\Pi^{12364}_{64}\sigma_{12364})\nonumber\\
+(Q^{45}_3dx^3)\cap(\Pi^{12345}_{45}\sigma_{12345}),
\end{align}
which in the isotropic case reduces to
\begin{align}
Q\cdot [\Pi_3]= -\bar{h}\left(\Pi^{2356}\sigma_{2356} + \Pi^{1436}\sigma_{1436}+\Pi^{1425}\sigma_{1425}\right).
\end{align}
Taking into account the spatial cycle $\Sigma_{xyz}$ we get that
\begin{align}
Q\cdot \left(\Sigma_{xyz}\times [\Pi_3]\right)= -\bar{h}\left(\Sigma_{xyz}\times \bigoplus_{i=1}^3 [\Pi_4]_i\right),
\end{align}
where $[\Pi_4]_i$ is the internal 4-cycle transversal to $T^2_i$ (e.g. $[\Pi_4]_2=\Pi^{1436}\sigma_{1436}$). This seven-cycle can be associated to the boundary of an 8-cycle in space-time, actually if
\begin{align}
[\Pi_8]=\left(\Sigma_{xyz}\times \bigoplus_{i=1}^3 [\Pi_4]_i\times \{-\infty, t_0]\right),
\end{align}
then $[\Pi_8]$ has a boundary at the time $t_0$. It follows that the cycle $[\Pi]=[\Pi_6]+\bar{h}[\Pi_8]$ satisfies  
$\partial_Q[\Pi]=0$ and hence it is a generalized $Q$-cycle. \\

Since the proposal is that D-branes would wrap only these kind of cycles, we get that an instantonic E5-brane wrapped on $[\Pi_3]$ supporting $-\bar{h}$ units of $Q$-fluxes $Q^{56}_1$, $Q^{64}_2$ and $Q^{45}_3$ would drive $\bar{h}$ $D7_i$-branes, for each $i$, to transform into flux $dF_1=-\bar{h}~dx^1\wedge dx^4$, which indeed corresponds to the magnetic field strength of the instantonic $E5$-brane. Notice that we have taken the convention in which the instantonic branes wrap the internal cycle in an orientation fixed by coordinates $\{x^i, x^{i+3}\}$ (with that ordering). Also we only made use of the non-geometric fluxes  in (\ref{bb}) and (\ref{hh}). The results for different configurations of instantonic branes are summarized in Tables \ref{tablaQ1}, \ref{tablaQ2} and \ref{tablaQ3}.\\


\begin{table}
\begin{center}
{\renewcommand{\arraystretch}{1.5}
\renewcommand{\tabcolsep}{0.2cm}
\begin{tabular}{|c|c|c|c|c|}
\hline
$E5$&$F_{abc}dx^{abc}$&$Q^{ab}_{c}$ for $D7_1$&$(Q\cdot F_3)_{14}$&RR flux\\
\hline\hline
456& $F_{123}\alpha_0$ & $Q^{23}_4=-h_1=-h$&$h$&$ -m\rightarrow -m+1$\\
\hline
123  & $F_{456}\beta_0$ & $Q^{56}_1=-\bar{h}_1=-\bar{h}$&$-\bar{h}$ &$-e_0\rightarrow -e_0+1$\\ 
\hline
423&$F_{156}\alpha_1$&$-Q^{56}_4=\bar{b}_{11}=\bar{\beta}$&$-\bar{\beta}$&$e\rightarrow e+1$\\
\hline
153 & $F_{426}\alpha_2$&$Q^{26}_1=\bar{b}_{12}=\bar{b}$&$\bar{b}$&$e\rightarrow e+1$\\
\hline
126 & $F_{453}\alpha_3$ & $Q^{53}_1=\bar{b}_{13}=\bar{b}$&$\bar{b}$ &$e\rightarrow e+1$\\
\hline
156 & $F_{423}\beta_1$ & $-Q^{23}_1=b_{11}=\beta$ &$\beta$& $-q \rightarrow -q+1$\\ 
\hline
426&$F_{153}\beta_2$&$Q^{53}_4=b_{12}=b$&$-b$&$ -q\rightarrow -q+1$\\
\hline
453& $ F_{126}\beta_3$&$Q^{26}_4=b_{13}=b$&$-b$&$-q\rightarrow -q+1$\\
\hline
\end{tabular}}
\caption{\small{Sources for the instability of $D7_1$-branes and the transformations on the corresponding RR fluxes. }}
\label{tablaQ1}
\end{center}
\end{table}

\begin{table}
\begin{center}
{\renewcommand{\arraystretch}{1.5}
\renewcommand{\tabcolsep}{0.2cm}
\begin{tabular}{|c|c|c|c|c|}
\hline
$E5$&$F_{abc}dx^{abc}$&$Q^{ab}_{c}$ for $D7_2$&$(Q\cdot F_3)_{25}$&RR flux\\
\hline\hline
456& $F_{123}\alpha_0$ & $Q^{31}_5=-h_2=-h$&$h$&$ -m\rightarrow -m+1$\\
\hline
123  & $F_{456}\beta_0$ & $Q^{64}_2=-\bar{h}_2=-\bar{h}$&$-\bar{h}$ &$-e_0\rightarrow -e_0+1$\\ 
\hline
423&$F_{156}\alpha_1$&$Q^{61}_2=\bar{b}_{12}=\bar{b}$&$\bar{b}$&$e\rightarrow e+1$\\
\hline
153 & $F_{426}\alpha_2$&$-Q^{64}_5=\bar{b}_{22}=\bar{\beta}$&$-\bar{\beta}$&$e\rightarrow e+1$\\
\hline
126 & $F_{453}\alpha_3$ & $Q^{34}_2=\bar{b}_{32}=\bar{b}$&$\bar{b}$ &$e\rightarrow e+1$\\
\hline
156 & $F_{423}\beta_1$ & $Q^{34}_5=b_{12}=b$ &$-b$& $-q \rightarrow -q+1$\\ 
\hline
426&$F_{153}\beta_2$&$-Q^{31}_2=b_{22}=\beta$&$\beta$&$ -q\rightarrow -q+1$\\
\hline
453& $ F_{126}\beta_3$&$Q^{61}_5=b_{32}=b$&$-b$&$-q\rightarrow -q+1$\\
\hline
\end{tabular}}
\caption{Sources for the instability of $D7_2$-branes and the transformations on the corresponding RR fluxes.}
\label{tablaQ2}
\end{center}
\end{table}

\begin{table}
\begin{center}
{\renewcommand{\arraystretch}{1.5}
\renewcommand{\tabcolsep}{0.2cm}
\begin{tabular}{|c|c|c|c|c|}
\hline
$E5$&$F_{abc}dx^{abc}$&$Q^{ab}_{c}$ for $D7_3$&$(Q\cdot F_3)_{36}$&RR flux\\
\hline\hline
456& $F_{123}\alpha_0$ & $Q^{12}_6=-h_3=-h$&$h$&$ -m\rightarrow -m+1$\\
\hline
123  & $F_{456}\beta_0$ & $Q^{45}_3=-\bar{h}_3=-\bar{h}$&$-\bar{h}$ &$-e_0\rightarrow -e_0+1$\\ 
\hline
423&$F_{156}\alpha_1$&$Q^{15}_3=\bar{b}_{13}=\bar{b}$&$\bar{b}$&$e\rightarrow e+1$\\
\hline
153 & $F_{426}\alpha_2$&$Q^{42}_3=\bar{b}_{23}=\bar{b}$&$\bar{b}$&$e\rightarrow e+1$\\
\hline
126 & $F_{453}\alpha_3$ & $-Q^{45}_6=\bar{b}_{33}=\bar{\beta}$&$-\bar{\beta}$ &$e\rightarrow e+1$\\
\hline
156 & $F_{423}\beta_1$ & $Q^{42}_6=b_{13}=b$ &$-b$& $-q \rightarrow -q+1$\\ 
\hline
426&$F_{153}\beta_2$&$Q^{15}_6=b_{23}=b$&$-b$&$ -q\rightarrow -q+1$\\
\hline
453& $ F_{126}\beta_3$&$-Q^{12}_3=b_{33}=\beta$&$\beta$&$-q\rightarrow -q+1$\\
\hline
\end{tabular}}
\caption{Sources for the instability of $D7_3$-branes and the transformations on the corresponding RR fluxes.}
\label{tablaQ3}
\end{center}
\end{table}

Observe as well that for each instantonic $E5$-brane wrapped on certain coordinates, there are 3 sources of non-geometric fluxes which drives 3 different D7-branes to become flux. Therefore although there are many D-branes transforming via the appearance of the instantonic brane, the increase on RR flux $F_3$ is still by one unit. Also notice there are 8 different ways in which a $D7$-brane transforms into flux, according to the non-geometric fluxes present in the configuration.\\

However,  there are restrictions  to turn on  any configuration of non-geometric fluxes we wish, since as shown in \cite{Aldazabal:2006up}, non-geometric fluxes must fulfill Bianchi constraints, namely that  $Q^{mn}_{[r}H_{ab]n}=Q^{[mn}_pQ^{r]p}_q=0$.  For isotropic fluxes, they read\footnote{Notice that in \cite{Aldazabal:2006up} there are 7 derived constraints. However it is easy to check that three of them are not independent. }
\begin{eqnarray}
\bar{h}_0h+\bar{a}(b+\beta)-a\bar{b}=0,\nonumber\\
h_0\bar{h}+a(\bar{b}+\bar{\beta})-\bar{a}b=0,\nonumber\\
h(\bar{b}+\bar{\beta})-b(b+\beta)=0,\nonumber\\
\bar{h}(b+\beta)-\bar{b}(\bar{b}+\bar{\beta})=0.
\label{bianchi}
\end{eqnarray}
It is easy to check that none of them suffer a change by a brane-flux transition mediated by 
a NS-NS-flux and an E5-brane. However, they will change if we allow S-dual transitions between RR and NS-NS-fluxes. In the presence of instantonic $NS5$-branes wrapping internal three-cycles, the units of the NS-NS flux in (\ref{HF1}) increases by one unit correspondingly. Let us say that there is one instantonic brane in each possible 3-cycle. Hence, in order to satisfy Bianchi identities, before and after the brane-flux transition, the non-geometric fluxes must fulfill the conditions
\begin{align}
h-(b+\beta)+\bar{b}=0\nonumber\\
\bar{h}-(\bar{b}+\bar{\beta})+b=0.
\label{Qbianchi}
\end{align}

Contrary to the absence of $Q$-fluxes, in this case brane-flux transitions are restricted. We observe that not all flux configurations fulfill the above constraints. Hence
the number of possible configurations giving rise to transitions are reduced.\\

\subsubsection{Vacuum solutions and brane-flux transitions}
Once we have studied how each $Q$-flux drives a specific brane-flux transition, we can go further and study the consequences on the supersymmetric  vacuum solutions in this isotropic non-geometric flux tori compactification. The superpotential related to the presence of non-geometric flux is \cite{Aldazabal:2006up}
\begin{align}
\W_Q(U,T)=\int (Q\cdot J_4(T))\wedge \Omega(U),
\end{align}
where the 4-form $J_4(T)$ is the complexified (quadratic) K\"ahler form which depends on the K\"ahler modulus $T$ as\footnote{We are taking the three K\"ahler moduli $T_i=T$.}
\begin{align}
J_4(T)= iT\sum_i \left(dx^{j}\wedge dx^{j+3}\wedge dx^{k}\wedge dx^{k+3}\right),
\end{align}
with cyclic coordinates $\{ i,j,k\}=\{1,2,3\}$. Using the expressions  (\ref{bb}), (\ref{hh}) and (\ref{omega}), we get
\begin{align}
\W_Q(U,T)=3T\left[-ih-U(2b+\beta)+iU^2(2\bar{b}+\bar{\beta}) +U^3\bar{h}\right]. 
\end{align}
Thus, in order to get a superpotential depending on all moduli, we must turn on NS-NS and RR 3-form fluxes as well, which together with the non-geometric fluxes, generate the superpotential $\W(S,\rho,\tau)$ given by
\begin{eqnarray}
\W(S,\rho,\tau)&=& \W_H(S,\rho)+\W_Q(\rho,\tau)= \int\left(F_3-SH_3+Q\cdot J_4\right)\wedge\Omega\nonumber\\
&=& (e_0+3e\rho+3q\rho^2-m\rho^3)-S(-h_0-3a\rho+3\bar{a}\rho^2+\bar{h}_0\rho^3)\nonumber\\
&&+3\tau\left[-h+\rho(2b+\beta)-\rho^2(2\bar{b}+\bar{\beta}) +\rho^3\bar{h}\right], 
\end{eqnarray}
with $\tau=iT$.  The superpotential is also written in terms of polynomials as $\W(S,\rho,\tau)=P_1(\rho)-SP_2(\rho)+\tau P_3(\rho)$, where
\begin{align}
P_3(\rho)= 3\left[-h+\rho(2b+\beta)-\rho^2(2\bar{b}+\bar{\beta}) +\rho^3\bar{h}\right].
\end{align}
The Minkowski vacuum solutions come from $\W=\partial_\rho\W=\partial_S\W=\partial_\tau\W=0$.
Under a brane-flux transition driven by non-geometric fluxes, only the polynomial $P_1(\rho)$ would suffer a change. As in the case without non-geometric fluxes, there are two cases. First, when the polynomial $P_1(\rho)$ just suffer a rescaling to $\kappa P_1(\rho)$. The three polynomials still share a common root $\rho_0$, but the values of the dilaton and the K\"ahler structure suffer a change,
\begin{align}
\partial_\rho P_1(\rho_0)-S\partial_\rho P_2(\rho_0)+\tau \partial_\rho P_3(\rho_0)=0\rightarrow \kappa\partial_\rho P_1(\rho_0)-\hat{S}\partial_\rho P_2(\rho_0)+\hat\tau \partial_\rho P_3(\rho_0)=0,
\end{align}
where one solution for the new moduli values $\hat{S}$ and $\hat\tau$ is $\hat{S}=\kappa S$ and $\hat\tau=\kappa\tau$. In this case, supersymmetry is still preserved and brane-flux transitions connect different supersymmetric vaccua. The second case concerns a change in the polynomial $P_1(\rho)$ to one which does not share a common factor with the other two polynomials. Supersymmetry is broken through all moduli in both points $\rho=\rho_0$ and $\rho=\rho_\ast$ being the latter the root of the transformed polynomial $\hat{P}_1(\rho)$.  A more detailed and extended analysys is required to see if the scalar potential has a minimum at $\rho=\rho_0$ or $\rho=\rho_\ast$ and evenmore  if it vanishes at those points.\\

However, the important thing here is that a non-geometric flux configuration is constrained to satisfy Bianchi identities (\ref{bianchi}), which restricts the values of the fluxes and thus  the  definition of  polynomials $P_2(\rho)$ and $P_3(\rho)$ with common complex roots.
In \cite{Aldazabal:2006up},  some classes of solutions satisfying these constraints were given.
 So, before considering brane-flux transitions, here we want to give another class of solutions for which, although there are common polynomials,  Minkowski SUSY vacuum solutions imply $Im(\rho)=0$. \\

\noindent
{\bf Example.}  Take the following set of relations,
\begin{eqnarray}
h_0=h= \pm b\qquad \qquad -3a=2b+\beta\nonumber\\
\bar{h}_0=\bar{h}=\pm\bar{b}, \qquad \qquad -3\bar{a}=2\bar{b}+\bar{\beta},
\label{U0}
\end{eqnarray}
with $\bar{a}=\pm a$, $\bar{b}=\pm b$ and $a=-b$.  Bianchi constraints are fullfiled for the two cases, upper and lower signs in the above relations.  For the first case,  the polynomials $P_2(\rho)$ and $P_3(\rho)$ are proportional,
\begin{align}
P_3(\rho)= 3P_2(\rho)= 3b(\rho^3-3\rho^2+3\rho -1)=3b(\rho-1)^3,
\end{align}
however, the solution for $\rho$ is real. For the second case, the relations between the polynomials read
\begin{align}
P_3(\rho)= 3P_2(\rho)= 3b(\rho^3+3\rho^2+3\rho +1)=3b(\rho+1)^3.
\end{align}

It happens that
within each of the two subclasses, the solutions for the complex structure $\rho$ are always real, implying that this set of solutions are located at the boundary of the moduli space. We can consider brane-flux transitions in order to move out from these solutions and try to get some complex solutions for $\rho$. For this to happen, $Q$-fluxes must satisfy the constraints in (\ref{Qbianchi}). Only the first case (upper signs) fulfill the constraints.\\

However,  brane-flux transitions mediated by instantonic RR charged branes, are going to modify only polynomial $P_1$ implying in some cases the breakdown of supersymmetry. Although a general analysis is beyond the goal of this paper, the simplest case involving a rescaling on the polynomial $P_1\rightarrow \kappa P_1$ involves the connection of different supersymmetric configurations with $Im(\rho)=0$. It would be very interesting to find a configuration in which the scalar potential is minimum and connected to the original one by some brane-flux transitions.  For this to succeed, the fluxes must also statisfy Bianchi identities and the constraints (\ref{Qbianchi}).\\

Since all these set of constraints make difficult to find interesting solutions, we rather take one step forward to include S-dual configurations. This work well for the case in which we only consider RR fluxes.
Following the prescription in \cite{Aldazabal:2006up} we move to S-dual versions in which extra fluxes induce new terms in the superpotential. \\

\subsection{S-dual vacuum transitions in non-geometric compactifications}
In type IIB 3-form flux compactification, D3-branes transform into flux in the presence of two different instantonic objects, namely E5-branes and NS5-branes. The latter one leaves as a remnant one unit of NS-NS flux $H_3$ which couples with $N$ units of the already present RR 3-form flux $F_3$, inducing $N$ units of  D3-brane charge. At the level of the field theory at the NS5-brane worldvolume, the presence of $F_3$ induces the coupling
\begin{align}
\int_{{\cal W}_{NS5}}F_3\wedge A_3.
\end{align}
The corresponding equations of motion are not fulfilled unless there are sources for $A_3$. This is what happens when a D3-brane ends at the 6-dimensional instantonic NS5-brane.\\

Notice here that the background NS-NS-flux $H_3$ is transversal to the instantonic NS5-brane. Hence a natural question involves a T-dual version of this system along two of the NS-NS-flux coordinates. We already know that the NS-NS-flux transforms into non-geometric flux $Q$ and since we are performing T-duality on coordinates transversal to the NS5-brane, this one will be invariant under the duality at least for the case in which the pull back of $F_3$ on the NS5-brane worldvolume vanishes.  Not being this the case, we should wonder what the T-dual version of a NS-NS-brane supporting RR 3-form flux is.\\

\subsubsection{S-dual non-geometric fluxes}
The presence of NS-NS, RR and non-geometric fluxes in type IIB tori compactification gives rise, as pointed out in \cite{Shelton:2006fd, Wecht:2007wu}, to a superpotential which is not S-dual.  The duality is recovered once we add extra fluxes as shown in \cite{Aldazabal:2006up}. These extra fluxes denoted as $P$ makes the superpotential S-dual and even more, close the corresponding algebra \cite{Aldazabal:2008zza, Font:2008vd}.  Although the origin of these fluxes has been not elucidated, their properties are easy to deduce. Being the S-dual version of $Q$, the S-dual non-geometric fluxes $P$ act on a $p$-form as
\begin{align}
P\cdot F_p= P^{ab}_cF_{ab\mu_1\dots\mu_{p-2}}dx^c\wedge dx^\mu_1\dots \wedge dx^{\mu_{p-2}},
\end{align}
while the action on a $p$-chain is
\begin{align}
P\cdot\Pi_p= P_{(ab)}\cap \Pi_{p(ab)},
\end{align}
similar as the $Q$ case. Hence, a $P$-cycle $[\Pi]$ (in the context of section 2) statisfies 
\begin{align}
\partial_P[\Pi]=\partial[\Pi]+P\cdot[\Pi]=0.
\end{align}

Using this definition is straightforward to check that a 5-brane supporting $P$-fluxes would drive some 7-branes to transform into fluxes. Since this configuration corresponds to the S-dual of a NS5-brane supporting  RR $F_3$ flux, we conclude that the involving 5-branes and 7-branes are nothing more than $NS5$-branes and 7-branes (S-dual to D7). These 7-branes have been well studied in literature \cite{Vafa:1996xn, Douglas:1996du, Sen:1996sk}. \\

Hence, we shall consider that a NS5-brane supporting $P$-flux is the S-dual version of a NS5-brane supporting RR $F_3$ flux\footnote{We leave for further work a detailed proof of such statement.}. This gives more insight about the nature of the S-dual non-geometric fluxes $P$. Using this, we see that $P$-fluxes also contribute to transitions between 7-branes and flux. In particular $P$-fluxes  give rise to an increase of NS-NS-flux units.\\

In order to study a concrete example, we shall take into account the $S$-dual configuration of $Q$-fluxes given in (\ref{bb}) and (\ref{hh}). Following notation in \cite{Aldazabal:2006up}, where essentially $Q$ is replaced  by $P$, the set of fluxes is

\begin{equation}
\mathbf{g}=\left(
\begin{array}{ccc}
-P^{23}_1 & P^{34}_5 & P^{42}_6\\
P^{53}_4 & -P^{31}_2 & P^{15}_6\\
P^{26}_4 & P^{61}_5 & -P^{12}_3
\end{array} \right)
\qquad 
\mathbf{\bar{g}}=\left(
\begin{array}{ccc}
-P^{56}_4 & P^{61}_2 & P^{15}_3\\
P^{26}_1 & -P^{64}_5 & P^{42}_3\\
P^{53}_1 & P^{34}_2 & -P^{45}_6
\end{array} \right),
\label{gg}
\end{equation}
and the two vectors
\begin{align}
\mathbf{-\overrightarrow{\mathbf{f}}}=\left( P^{23}_4, P^{31}_5, P^{12}_6\right) \qquad -\mathbf{\overrightarrow{\mathbf{\bar{f}}}}=\left(P^{56}_1, P^{64}_2, P^{45}_3\right).
\label{ff}
\end{align}
In the isotropic case,  $\mathbf{g}_{ij}=g$ for $i\neq j$ and $\mathbf{g}_{ii}=\gamma$ with similar expressions for $\mathbf{\bar{g}}$ and $\mathbf{\bar{g}}$\\

The presence of  $g$ and $\gamma$  $P$-fluxes drive some 7-branes to decay into a new configuration of fluxes, by increasing the NS-NS units. The relation between the units of $P$-flux and the corresponding NS-NS-flux increments are summarized in tables \ref{tablaP1}, \ref{tablaP2} and \ref{tablaP3}.\\

\begin{table}
\begin{center}
{\renewcommand{\arraystretch}{1.5}
\renewcommand{\tabcolsep}{0.2cm}
\begin{tabular}{|c|c|c|c|c|}
\hline
$NS5$&$H_{abc}dx^{abc}$&$P^{ab}_{c}$ for $7_1$&$(P\cdot H_3)_{14}$&NS-NS flux\\
\hline\hline
456& $H_{123}\alpha_0$ &   $P^{23}_4=-f_1=-f$&$f$&$ \bar{h}_0\rightarrow \bar{h}_0+1$\\
\hline
123  & $H_{456}\beta_0$ &   $P^{56}_1=-\bar{f}_1=-\bar{f}$&$-\bar{f}$ &$h_0\rightarrow h_0+1$\\ 
\hline
423&$H_{156}\alpha_1$&     $-P^{56}_4=\bar{g}_{11}=\bar{\gamma}$&$-\bar{\gamma}$&$-a\rightarrow -a+1$\\
\hline
153 & $H_{426}\alpha_2$&   $P^{26}_1=\bar{g}_{12}=\bar{g}$&$\bar{g}$&$-a\rightarrow -a+1$\\
\hline
126 & $H_{453}\alpha_3$ &  $P^{53}_1=\bar{g}_{13}=\bar{g}$&$\bar{g}$ &$-a\rightarrow -a+1$\\
\hline
156 & $H_{423}\beta_1$ & $-P^{23}_1=g_{11}=\gamma$ &$\gamma$& $-\bar{a} \rightarrow -\bar{a}+1$\\ 
\hline
426&$H_{153}\beta_2$&$P^{53}_4=g_{12}=g$&$-g$&$ -\bar{a}\rightarrow -\bar{a}+1$\\
\hline
453& $ H_{126}\beta_3$&$P^{26}_4=g_{13}=g$&$-g$&$-\bar{a}\rightarrow -\bar{a}+1$\\
\hline
\end{tabular}}
\caption{\small{Sources for the instability of $7_1$-branes and the transformations on the corresponding NS-NS fluxes.}}
\label{tablaP1}
\end{center}
\end{table}

\begin{table}
\begin{center}
{\renewcommand{\arraystretch}{1.5}
\renewcommand{\tabcolsep}{0.2cm}
\begin{tabular}{|c|c|c|c|c|}
\hline
$NS5$&$H_{abc}dx^{abc}$&$P^{ab}_{c}$ for $7_2$&$(P\cdot H_3)_{25}$&NS-NS flux\\
\hline\hline
456& $H_{123}\alpha_0$ & $P^{31}_5=-f_2=-f$&$f$&$ \bar{h}_0\rightarrow \bar{h}_0+1$\\
\hline
123  & $H_{456}\beta_0$ & $P^{64}_2=-\bar{f}_2=-\bar{f}$&$-\bar{f}$ &$h_0\rightarrow h_0+1$\\ 
\hline
423&$H_{156}\alpha_1$&$P^{61}_2=\bar{g}_{12}=\bar{g}$&$\bar{g}$&$-a\rightarrow -a+1$\\
\hline
153 & $H_{426}\alpha_2$&$-P^{64}_5=\bar{g}_{22}=\bar{\gamma}$&$-\bar{\gamma}$&$-a\rightarrow -a+1$\\
\hline
126 & $H_{453}\alpha_3$ & $P^{34}_2=\bar{g}_{32}=\bar{g}$&$\bar{g}$ &$-a\rightarrow -a+1$\\
\hline
156 & $H_{423}\beta_1$ & $P^{34}_5=g_{12}=g$ &$-g$& $-\bar{a} \rightarrow -\bar{a}+1$\\ 
\hline
426&$H_{153}\beta_2$&$-P^{31}_2=g_{22}=\gamma$&$\gamma$&$ -\bar{a}\rightarrow -\bar{a}+1$\\
\hline
453& $ H_{126}\beta_3$&$P^{61}_5=g_{32}=g$&$-g$&$-\bar{a}\rightarrow -\bar{a}+1$\\
\hline
\end{tabular}}
\caption{Sources for the instability of $7_2$-branes and the transformations on the corresponding NS-NS fluxes.}
\label{tablaP2}
\end{center}
\end{table}

\begin{table}
\begin{center}
{\renewcommand{\arraystretch}{1.5}
\renewcommand{\tabcolsep}{0.2cm}
\begin{tabular}{|c|c|c|c|c|}
\hline
$NS5$&$H_{abc}dx^{abc}$&$P^{ab}_{c}$ for $7_3$&$(P\cdot H_3)_{36}$&NS-NS flux\\
\hline\hline
456& $H_{123}\alpha_0$ & $P^{12}_6=-f_3=-f$&$f$&$ \bar{h}_0\rightarrow \bar{h}_0+1$\\
\hline
123  & $H_{456}\beta_0$ & $P^{45}_3=-\bar{f}_3=-\bar{f}$&$-\bar{f}$ &$h_0\rightarrow h_0+1$\\ 
\hline
423&$H_{156}\alpha_1$&$P^{15}_3=\bar{g}_{13}=\bar{g}$&$\bar{g}$&$-a\rightarrow -a+1$\\
\hline
153 & $H_{426}\alpha_2$&$P^{42}_3=\bar{g}_{23}=\bar{g}$&$\bar{g}$&$-a\rightarrow -a+1$\\
\hline
126 & $H_{453}\alpha_3$ & $-P^{45}_6=\bar{g}_{33}=\bar{\gamma}$&$-\bar{\gamma}$ &$-a\rightarrow -a+1$\\
\hline
156 & $H_{423}\beta_1$ & $P^{42}_6=g_{13}=g$ &$-g$& $-\bar{a} \rightarrow -\bar{a}+1$\\ 
\hline
426&$H_{153}\beta_2$&$P^{15}_6=g_{23}=g$&$-g$&$ -\bar{a}\rightarrow -\bar{a}+1$\\
\hline
453& $ H_{126}\beta_3$&$-P^{12}_3=g_{33}=\gamma$&$\gamma$&$-\bar{a}\rightarrow -\bar{a}+1$\\
\hline
\end{tabular}}
\caption{Sources for the instability of $7_3$-branes and the transformations on the corresponding NS-NS fluxes.}
\label{tablaP3}
\end{center}
\end{table}

Before analyzing the vacuum transitions, it is important to stress out that, as for $Q$-fluxes, $P$-fluxes also satisfy Bianchi identities given by $P\cdot P= P^{[MN}_SP^{L]P}_R=0$ and $Q\cdot P+P\cdot Q= Q^{[MN}_SP^{L]P}_R+P^{[MN}_SQ^{L]P}_R=0$. These identities are invariant under transitions driven by instantonic $NS5$-branes if, altogether with constraints (\ref{Qbianchi}), $P$-fluxes fulfill the following relations,
\begin{align}
f-(g+\gamma)+\bar{g}=0\nonumber\\
\bar{f}-(\bar{g}+\bar{\gamma})+g=0.
\label{PBianchi}
\end{align}

In this case, the superpotential is given by \cite{Aldazabal:2006up},
\begin{align}
\W(S, \rho, \tau)=\int_{T^6}\left(F_3-SH_3+(Q-SP)\cdot J_c\right)\wedge \Omega,
\end{align}
which in terms of polynomials read
\begin{align}
\W(S, \rho, \tau)= P_1(\rho)-S P_2(\rho)+ \tau P_3(\rho) - S\tau P_4(\rho),
\end{align}
with 
\begin{align}
P_4(\rho)= -3\left(f-(2g+\gamma )\rho +(2\bar{g}+\bar{\gamma} )\rho^2-\bar{f}\rho^3\right),
\end{align}
and $P_1(\rho)$ , $P_2(\rho)$ and $P_3(\rho)$ given as before. SUSY solutions imply that
\begin{align}
\tau=\frac{P_2}{P_4}, \qquad S=\frac{P_3}{P_4},
\end{align}
and the complex structure is determined as the common root of 
\begin{align}
P_1P_4-P_2P_3=0\nonumber\\
(P_4P_1'-P_1P_4')+(P_2P_3'-P_3P_2')=0.
\end{align}

\subsubsection{Vacuum transitions}
Finding general Minkowski solutions in S-dual non-geometric tori compactifications is quite complicated since there are several constraints to fulfill.  Thus, we shall present some general  aspects about the effects of brane-flux transition on the superpotential and the corresponding Minkowski SUSY vacuum solutions and later on, we shall study some specific examples from which we can get some important conclusions.\\

First of all notice that only polynomials $P_1(\rho)$ and $P_2(\rho)$ will be affected by a brane-flux transition. Thus in this case, brane-flux transitions will modify the value of $\tau$ and the complex structure $\rho$, keeping the dilaton $S$ invariant.  For this to happen, it is also required that Bianchi identities be invariant as well.  In case we find a vacuum on which all these requirements are satisfied,  transitions between branes and fluxes would connect two different vacua with the same dilaton value.  Let us concentrate on specific examples.\\

\noindent
{\bf Example 1.}  In \cite{Aldazabal:2006up} two classes of solutions were studied in which Bianchi identities were fulfilled and viable Minkowski SUSY solutions were found. Taking the set of relations,
\begin{align}
f=\bar{f}=g=\bar{g}=0, \quad \beta=-b, \quad \bar\beta=-\bar{b}, \quad b\gamma=h\bar\gamma, \quad h_0\bar{b}=e\gamma +\bar{a}h, \quad ,\nonumber\\
\bar{h}_0h=q\gamma +a \bar{b}, \quad q=0, \quad e_0\gamma=4ah,\quad mh_0h\gamma=(e\gamma+4h\bar{a})(e\gamma+h\bar{a}),\nonumber
\end{align}
there is a Minkowski SUSY vacuum with 
\begin{align}
S=\frac{2h}{\gamma U}, \quad T=\frac{h(h_0(2\bar{a}h-e\gamma)-2iaU(e\gamma+\bar{a}h))}{3\gamma U(e\gamma+\bar{a}h)(h-ibU)}\nonumber\\
U=(-\frac{h_oh}{e\gamma+h\bar{a}})^{1/2}.\nonumber
\end{align}
However, this solution does not change via the appearance of instantonic branes $E5$ or $NS5$, since the constraints (\ref{Qbianchi}) and (\ref{PBianchi}) are not fullfilled by these set of fluxes. Actually, in this case constraints to keep Bianchi identities imply that $\gamma=\bar\gamma=0$, which tells us that the only compatible condition for the $P$-fluxes is to vanish.\\

The same happens with the second set of fluxes studied in \cite{Aldazabal:2006up}, which take as  solutions to Bianchi identities 
\begin{align}
\gamma=-g, \quad \bar\gamma=-\bar{g}, \quad, f\bar{f}=g\bar{g},\quad{f}\bar\beta=g\beta,\nonumber\\
h=\bar{h}=b=\bar{b}=0,\quad mg=e\bar{f}-\bar{a}\bar\beta,\quad qg=\alpha\bar\beta-e_0\bar{f},\nonumber
\end{align}
and Minkowski vacuum solutions are guaranteed if
\begin{align}
\bar{a}=0,\quad h_0\bar\beta=4eg, \quad \bar{h}_0\bar\beta=\frac{\bar{f}(4e_0\bar{f}-3a\bar\beta)}{g}.\nonumber
\end{align}
The solutions for the moduli are nevertheless protected to suffer transformations by brane-flux transitions, since the set of fluxes do not fulfill the constraints to keep the Bianchi identities. \\

\noindent
{\bf Example 2.}  Consider the set of fluxes in (\ref{U0}) and
\begin{eqnarray}
-e_0=f=\pm g, \quad 3e=(2g+\gamma), \nonumber\\
-m=\bar{f}=\pm \bar{g}\quad -3q=(2\bar{g}+\bar\gamma), 
\end{eqnarray}
with $q=\mp e$, $g=\pm\bar{g}$ and $ g=e$, for which the Bianchi identities are fulfilled. The constraints for the Bianchi identities to be satisfied after a transition are fulfilled just for the case in which we take the upper signs on the above set of conditions. In that case, we can take an extra relation as $a=-b=g=e$. This allows us to construct the four polynomials as
  \begin{align}
-3P_1=3P_2=P_3=-P_4= 3b(-1+3\rho-3\rho^2+\rho^3),
\end{align}
which give a real solution for $\rho$.  Among this set of fluxes, we see that solutions with such a complex structure are connected through the presence of instantonic branes to other ones. Notice however, that those solutions are connected to the boundary of the moduli space, and are not interesting for phenomenological purposes.\\

It seems, at least from these examples,  that viable solutions (with complex $\rho$) are protected to move by  instantonic brane mediations. On the other hand, solutions in the boundary of moduli space (Im $\rho=0$) are allowed to move via the brane-flux transitions.


\section{Final comments and conclusions}
Incorporation of fluxes in string compactifications has enriched our notion, not only about stabilization of  fluxes, but also about the geometry of the space on which strings and D-branes are defined. A particular interesting case involves the mathematical definition of a cycle on which a D-brane can be safely wrapped (i.e., cancellation of Freed-Witten anomalies ).  Instead of canceling Freed-Witten anomalies by vanishing the pullback of fluxes on the D-brane's worldvolume, we consider the cancellation by adding extra branes.  In this paper we have extended the notion of twisted homology developed in \cite{Collinucci:2006ug} for the case in which geometric and non-geometric fluxes are present. In such context, a generalized cycle can be thought as the locus on which D-branes wrap, rendering the system free of anomalies (induced by the fluxes). However, instead of having single D-branes wrapping cycles, we have a net of D-branes (consisting on D-branes which end at another one  with different dimension) wrapping a linear combination of chains, referred to as twisted cycles .\\

Hence, the notion of boundary is generalized according to the fluxes present in the system. For instance, in the case of non-geometric background, the generalized boundary reads $\partial_Q = \partial + Q\cdot$,
where $Q\cdot$ indicates the action of non-geometric fluxes on a chain through out the cap product (section 2.3). In this context, a generalized $Q$-cycle is a linear combination of cyclles and chains. As an example, we found that in the presence of the flux $Q^{12}_3=N$,  one generalized cycle reads
$[\Pi]=[\Sigma_1]-N\Sigma_3$,
with $\partial \Sigma_3\neq 0$. This tells us that an anomaly free system consists on a $D1$-brane wrapping the spatial cycle $[\Sigma_1]$ in Minkowski space and a $D3$-brane wrapping the internal $\Sigma_3$. The fact that the metric on a torus threaded with non-geometric flux is not globally defined, follows from the result that $\Sigma_1$ is not a generalized cycle to be wrapped by a D-brane. Hence all 1-forms with support on it will be ill-defined.  By adding extra branes, the anomaly is cured and we can overcome the initial problem of wrapping D-branes along a one-cycle.\\

In the second part of the paper we have concentrated on the above results for the case in which the anomalous brane is localized in time, i.e., instantonic. In such situation, for each unit of background flux, a D-brane becomes unstable to decay into a configuration of fluxes, which involves the already background ones and the magnetic field strength related to the instantonic brane. We call this process a brane-flux transition.\\

Now,  it is known that cancellation of Freed-Witten anomalies by enforcing the flux value at the worldvolume to vanish,  translates into invariance of the superpotential $\W$. In our case the superpotential $\W$ is not invariant implying the presence of connections between different Minkowski vacuum solutions from F-flat conditions.\\

Although brane-flux transitions leave the tadpole condition invariant (after all, tadpole counts the total number of D-brane charge from fluxes and from physical D-branes), they perform a change in  $\W$ which depends on RR fluxes. This essentially follows, as mentioned, from the fact that there is a remnant  of one unit of RR flux attributed to the instantonic brane. By considering the S-dual case as well,  brane-flux transition is driven also by instantonic NS5-branes.\\

Our main goal was to study the effects of brane-flux transitions on the vacuum solutions of type IIB flux compactifications on a factorizable six-dimensional torus in a simple orientifold plane. Vacuum  solutions can be written in terms of polynomials $P_1$ and $P_2$ which share a common factor $P$. All polynomials depend on the complex structure $U=i\rho$ and on the units of RR and NS-NS flux. Summarizing,  before the transition, complex structure $\rho$ is stabilized at the commun root $\rho_0$ of $P(\rho)$, and the dilaton $S(\rho)$ fulfills the equation $P_1'(\rho)-SP_2'(\rho)=0$. After the transition we have three general cases:
\begin{enumerate}
\item
$P_1(\rho)=P(\rho)Q_1(\rho)\rightarrow \hat{P}_1(\rho)=P(\rho)\hat{Q}_1(\tilde{f}_1\rho+\tilde{g}_1)$, where $\tilde{f}$ and $\tilde{g}$ are functions of the modified RR fluxes. The complex structure is invariant (there is still a common root between $P_2$ and $\hat{P}_1$), but the value at which the dilaton is stabilized suffers a change. The prototypical example for this case takes a transformation on $P_1(\rho)$ to $\kappa P_1(\rho)$ from which one gets that the complex structure is stabilized at the same value $\rho_0$.  However, from the F-flat equation $\partial_\rho\W_H=0$ we get that the dilaton $S$ must fullfill the equation
$\kappa P_1'(\rho)- S P_2'(\rho)=0$,
which is true for $S=\kappa \rho_0$.
\item
The transition changes the polynomial such that there is  non-common factor $P(\rho)$. The new polynomial  reads $\hat{P}_1(\rho)=\hat{P}(\rho)\hat{Q}_1(\rho)$ with $\hat{P}(\rho) =0$ at $\rho=\rho_*(\neq \rho_0)$. Hence, at $\rho=\rho_\ast$  and at $\rho=\rho_0$, $\W_H\neq 0$ from which we see that supersymmetry is broken through all moduli. At both points, the scalar potential is positive (since the model is no-scale) but an extended and more detailed analysis is required to elucidate if the scalar potential is minimum. In the case in which the scalar potential is not minimum at those points, we can say that brane-flux transitions connect the supersymmetric configuration into some unstable excitations around the basis point.  
\item
By considering transitions mediated by NS5-branes, 
we find an example in which the transition changes the polynomial $P_1(\rho)$ and $P_2(\rho)$  such that there is still a common root since they share a common factor.  However, this common factor $Q(\rho)$ has also changed with respect to $P(\rho)$ gathered from the original set of fluxes. The new polynomials are  $P_1(\rho)= Q(\rho)(\tilde{f}_1\rho +\tilde{g}_1)$ and $P_2(\rho)=Q(\rho)(\tilde{f}_2\rho + \tilde{g}_2)$ with $Q(\rho_\ast)=0$ and $\rho_\ast\neq\rho_0$. In this case, the transition connects two different supersymmetric solutions which differ not only by a rescaling of fluxes. Therefore, the nucleation of instantonic NS5-branes 
allows us to connect many different vacuum solutions which otherwise would be disconnected. The S-dual version of the brane-flux transition must  notably reduce the size of vacuum solutions.\\
\end{enumerate}

These results do not involve the presence of non-geometric fluxes. Their incorporation establishes novel ways to increase the value of RR and NS-NS fluxes.  An instantonic  $E5$-brane supporting non-geometric fluxes in its worldvolume would drive some D3-brane to transform into flux. However, non-geometric fluxes, as well as their S-duals are constrained to satisfy Bianchi identities. Since a general configuration of fluxes satisfying all the required constraints is quite difficult to obtain, we have concentrated our study to some very specific examples from which we conclude the following:
\begin{enumerate}
\item
For the cases in which the solution to SUSY equations allows a non vanishing complex structure $U$ with $Re(U)>0$, the configuration of fluxes does not satisfy the required constraints for the transition to happen. This mean that brane-flux transition is forbidden for these cases, isolating the vacuum solution to connect to others. Notice that these solutions are interesting from the phenomenology point of view and are protected to move to another vacuum via an instantonic brane mediation.
\item
For the cases in which we get a vanishing complex structure, and hence, a non-interesting configuration, the set of fluxes satisfy the constraints which allow the transition to occur. Different vacuum solutions, all of them sharing the property that $Re(U)=0$, are connected through a chain of  instantonic branes.
\end{enumerate}

It is clear that a deeper  study on different vacuum solutions is required. We leave it for further study. However, it is very interesting to notice some common results from this work and reference \cite{Font:2008vd} in which the authors study the non-geometric fluxes algebra and some relations between different Minkowski and AdS$_4$ vacuum solutions.

\vspace{1cm}

\begin{center}
{\bf Acknowledgments}
\end{center}

We thank Hugo Garc\'ia Compe\'an for support and for discussions. Also we are grateful to  Andrei Micu for enlighten  explanations. O. L.-B. thanks Alexandra Delgado for kind support and encouragement. W. H.-S. thanks CINVESTAV-Monterrey for kind hospitality. O. L-B. was partially supported by CONACyT-60209  ``Topological effects on string phenomenology".

\appendix

\section{Cap product}

In this appendix we would like to briefly review some basic notions on cycles, simplexes and cap product. At the end, we show an explicit calculation of the cap product between a form and a cycle. The main references for this appendix are \cite{Hatcher:2002ht} and \cite{Collinucci:2006ug}.\\

 \subsection{Definitions}
 Here we will define the cap product between homology and
 cohomology. 
 Let $X$ be a topological space. A \textbf{singular p-simplex} is
a continuous map $ \sigma : \Delta^p \rightarrow X$, where $\Delta
^p$ is the \textbf{standard p-simplex}:
\begin{equation}
    \Delta^p \equiv\big{\{} \Sigma _{i=0}t_i v_i :  \ 0 \leq t_i \leq 1 \ \hbox{and} \
    \Sigma_{i=0}^{k}=1 \big{\}}.
    \label{psimplex}
 \end{equation}
 where $v_0=0$ and $v_i$ is the $i$th standard basis vector. It is
important to remark that there is an ordering on the vertices of
the p-simplex defined by the integer that identifies each of them.
This ordering determines an orientation according to increasing
subscripts.\\

 Let $C_p(X)$ be the free abelian group generated by the set of
all $p$-simplices in $X$. An element of this group is called a
\textbf{singular p-chain} and can be visualized as a finite linear
combination of singular p-simpleces with integer coefficients. It
can also be defined a boundary map $\partial:C_p(X) \rightarrow
C_{p-1}(X)$:
\begin{equation}
\partial(\sigma)=\Sigma_i (-1)^i \sigma \mid \Delta^{p-1}_i,
 \label{bmap}
\end{equation}
where $\sigma \mid \Delta^{p-1}_i$ denotes the restriction of
$\sigma$ to the standard $(p-1)$-simplex $\Delta^{p-1}_i$ spanned by
all basis vectors spanning $\Delta ^p$ but $v_i$.
It can be shown that $\partial ^2=0$, so we can define the
\textbf{singular homology group} $H_p(X)=Ker \partial /
Im{\partial}$. Elements of Ker$\partial$ are called
\textbf{cycles} and elements of Im$\partial$ are called
\textbf{boundaries}.\\

By identifying each singular simplex $\sigma$ with its image
$\sigma(\Delta ^p) \subset X$, we can associate to each
$p$-simplex a smooth $p$-submanifold of $X$. Hence a p-chain can
be seen as a formal sum of p-submanifolds of $X$, weighed by
integers. Thus we can consider singular homology as a
classification of submanifolds and by abuse of language we will
also call cycles to submanifolds corresponding to cycles in
singular homology. The manifolds that D-branes can wrap are
precisely these cycles; and the weights are the charges of the
D-branes that wrap them.\\

Let us now review the case of 
singular cohomology with integer coefficients.
Given a space $X$, we define the group $C^{p}(X)$ of
\textbf{singular p-cochains with coefficients in $\mathbb{Z}$} to
be the dual group Hom($C_p(X);\mathbb{Z}$) of the singular chain
group $C_p(X)$. In particular a $p$-cochain $\varphi \in C^p(X)$
assigns to each singular p-simplex $\sigma : \Delta^p \rightarrow
X$ an integer $\varphi(\sigma)$.\\

We can also define the \textbf{coboundary map} $\delta : C^p(X)
\rightarrow C^{p+1}(X)$ as the dual map to the chain boundary map
$\partial$. This means that for a singular p+1 simplex $\sigma :
\Delta ^{p+1} \rightarrow X$ we have\\
\begin{equation}
\delta \varphi (\sigma) = \Sigma_i (-1)^i \varphi (\sigma \mid
\Delta^{p}_i \label{cbmap}),
\end{equation}


It is an easy fact that $\delta ^2 =0.$ Hence we can define the
\textbf{singular cohomology group with coefficients in
$\mathbb{Z}$}: $H^n(X)=Ker \delta / Im \delta.$ Elements of
Ker$\delta$ are called \textbf{cocycles} and elements of
Im$\delta$ are called \textbf{coboundaries}.\\


We have all the requirements to define a product between co-chains and chains, through out the cap product. This is as follows.
Let $X$ be a topological space. An $\mathbb{Z}$-bilinear
map (called the cap product) is defined as  $\cap :C^{l}(X) \times C_k(X)
\rightarrow C_{k-l}(X)$ for $k \geq l$ by setting
\begin{equation}
\sigma \cap \varphi = \varphi (\sigma \mid \Delta^{l}_{[v_0,
\ldots, v_l]}) \ \sigma \mid \Delta ^{k-l}_{[v_l, \ldots, v_k]},
 \label{cap}
\end{equation}
for $\sigma : \Delta^k \rightarrow X$ and $\varphi \in C^l(X)$ (and
the corresponding extensions by linearity). In (\ref{cap}), $\sigma \mid
\Delta^{l}_{[v_0, \ldots, v_l]}$ denotes the $l$-simplex spanned
by  $v_0$ (the zero vector) and the first $l$ vectors $v_l$ that
span the standard k-simplex $\Delta_k$ and $\Delta ^{k-l}_{[v_l,
\ldots, v_k]}$ denotes the standard $(k-l)$-simplex spanned by
$v_l, \ldots, v_k \footnote{Strictly speaking, $v_l, \ldots, v_k$
does not expand a standard $k-l$ simplex because $v_l$ is not the
cero vector. However, it is easy to see that the simplex spanned
by these vectors can be maped homeomorphically to the standard
$(k-l)$-simplex }.$ It turns out that the cap product is well
defined as a map $H^l(X) \times H_k(X) \rightarrow H_{k-l}(X)$;
but in this paper our interest on it will be as a map $H^l(X)
\times C_k(X) \rightarrow C_{k-l}(X)$.\\

For application in the context of this article,  the interpretation of the cap product one should have in mind is
as follows. Suppose there is a D$p$-brane in type II superstring
theory  and also there is a nontrivial $H$-flux in the bulk such
that the $p$-cycle wrapped by the D$p$-brane is Freed-Witten
anomalous. Using the de Rham theorem, we can identify the
class $[H]$ defined by the $H$-flux in the third group of de Rham
cohomology with a class (also denoted $[H]$) in the third singular
cohomology group $H^3(X)$. Then the cap product of $[H]$ with the
$p$-chain wrapped by the D$p$-brane produces a $(p-3)$-chain
representing the codimension 3 submanifold (with respect the
original D$p$-worldvolume) of the magnetic monopole in the
worldvolume $U(1)$ gauge theory of the D$p$-brane. This magnetic
monopole is necessary for cancelling the Freed-Witten anomaly and
thus rendering this system consistent \cite{Collinucci:2006ug}.\\

\subsection{Applications}
From the above paragraph it should be clear that for our
applications to D-brane physics is desirable to have a description
of the cap product in terms of differential forms. Here we
describe this approach.\\

 For the
purpose of this section it is more convenient to denote a
$k$-cycle, in analogy with the usual notation for differential
forms, as $[\Pi_k]= C^{\mu_1 \ldots \mu_k} \sigma_{\mu_1 \ldots
\mu_k}$, where $C^{\mu_1 \ldots \mu_k}$ are integers and
$\sigma_{\mu_1 \ldots \mu_k}$ is a basis for a $k$-chain constructed from $k$-simplexes. In this language,
the cap product between an $l$-form and a $k$-dimensional
submanifold (remember that a $k$-chain can be identified with a
$k$-dimensional submanifold) $[\Pi_k]$ where a brane is wrapped,
is the $k-l$ submanifold  which is Poincar\'e dual to the pullback
of the $l$-form into the brane worldvolume. This is a codimension
$k$-submanifold in the worldvolume.\\

The precise form of the cap product is as follows: consider the
l-form 
\begin{align}
f(x^{i_1}, \ldots, x^{i_l})dx^{i_1 \ldots i_l} \equiv
f(x^{i_1}, \ldots, x^{i_l}) dx^{i_1} \wedge dx^{i_2} \wedge\ldots
 \wedge dx^{i_l},
 \end{align}
  and the $k$-cycle $[\Pi_k]=
C^{\mu_1 \ldots \mu_k} \sigma_{\mu_1 \ldots \mu_k}.$ Then, their
the cap product is
\begin{eqnarray}
dx^{i_1 \ldots i_l} \cap [\Pi_k] &=& C^{\mu_1 \ldots \mu_k} dx^{i_1
\ldots i_l}  \cap \sigma_{\mu_1 \ldots \mu_k}\\
&=& C^{\mu_1 \ldots \mu_k} \Sigma_{l \{\mu_1 \ldots \mu_k\}}
\big{(}\int _{\sigma _{l \{\mu_1 \ldots \mu_k\} }} f(x^{i_1},
\ldots, x^{i_l}) dx^{i_1 \ldots i_l} \big{)} \hat{\sigma}_{l
\{\mu_1 \ldots \mu_k\}},\nonumber
 \label{capdifferential}
\end{eqnarray}
where $l \{\mu_1 \ldots \mu_k\}$ are the $l$-uplas formed by
picking $l$ indices from the set $ \{\mu_k \}$; $\sigma _{l
\{\mu_1 \ldots \mu_k\}}$ are the $l$-cycles obtained from
$\sigma_{\mu_1 \ldots \mu_k}$ by picking out just the indices that are
contained in $l \{\mu_1 \ldots \mu_k\}$ and $\hat{\sigma}_{l
\{\mu_1 \ldots \mu_k\}}$ is the $k-l$-cycle obtained from
$\sigma_{\mu_1 \ldots \mu_k}$ by throwing out the $l$-upla $l
\{\mu_1 \ldots \mu_k\}.$ The product with more general $l$-forms
is obtained extending by linearity.\\

Let us now illustrate, by an example described in the body of this
paper, how to carry over the cap product. For this we will
identify (as we have just mentioned in the above paragraph) the
cohomology class twisting the ordinary homology with a three-form
in de Rham cohomology, which physically is the field strength of
the NS-NS B-field.\\

Consider the six-torus  $T^6=(T^2)^3$ threaded with a NS-NS-flux.
Take as coordinates for $T^6$ the pairs $(x^{i},x^{i+3})$ with
$i=1,2,3$ where  the pair $(x^{i},x^{i+3})$ parametrizes the
$i$th-torus.  For definiteness we will consider the NS-NS-flux
$H_3=Ndx^1 \wedge dx^2 \wedge dx^3,$ with $N$ an integer. Let us
also take as basis for the one-cycles the pairs $([a_i],[b_i])$ with
$\int_{[a_i]}dx^j = \int_{[b_i]}dx^{j+3} = \delta_i ^j$ and
$\int_{[a_i]}dx^{j+3} = \int_{[b_i]}dx^j=0.$\\

Consider now the 3-cycle $[\Pi_3]=\otimes_{i=1}^3 n^{i}[a_i],$
with the $n^i$ integers. We want to calculate the cap product $H
\cap [\Pi_3].$
\begin{equation}
H \cap [\Pi_3]= Nn^1 n^2 n^3 (dx^1 \wedge dx^2 \wedge dx^3) \cap
([a^1] \times [a^2] \times [a^3]) \label{capexample}
\end{equation}

In terms of differential forms, the evaluation of a cohomology
class in a homology class is substituted by the standard
integration of the corresponding differential form on the
corresponding cycle. Hence we have

$$ H \cap [\Pi_3]= [0]Nn^1 n^2 n^3 \int_{[a^1] \times [a^2] \times
[a^3]}(dx^1 \wedge dx^2 \wedge dx^3)$$

\begin{equation}
 \qquad \qquad \quad = [0]\int_{[a_1]}dx^1
\int_{[a_2]}dx^2 \int_{[a_3]}dx^3 = [0]Nn^1 n^2 n^3,
 \label{capexample2}
\end{equation}
where $ [0] \in H^{0}(T^6)$ is the zero cycle generating
$H^{0}(T^6) = \mathbb{Z}.$

\bibliography{Fwng}

\providecommand{\bysame}{\leavevmode\hbox to3em{\hrulefill}\thinspace}
\begin{thebibliography}{10}

\bibitem{Grana:2005jc}
M.~Grana, \emph{{Flux compactifications in string theory: A comprehensive
  review}}, Phys. Rept. \textbf{423} (2006), 91--158,  \texttt{hep-th/0509003}.

\bibitem{Douglas:2006es}
M.~R. Douglas and S.~Kachru, \emph{{Flux compactification}}, Rev. Mod. Phys.
  \textbf{79} (2007), 733--796,  \texttt{hep-th/0610102}.

\bibitem{Shelton:2005cf}
J.~Shelton, W.~Taylor, and B.~Wecht, \emph{{Nongeometric Flux
  Compactifications}}, JHEP \textbf{10} (2005), 085,  \texttt{hep-th/0508133}.

\bibitem{Wecht:2007wu}
B.~Wecht, \emph{{Lectures on Nongeometric Flux Compactifications}}, Class.
  Quant. Grav. \textbf{24} (2007), S773--S794,  \texttt{0708.3984}.

\bibitem{Aldazabal:2008zza}
G.~Aldazabal, P.~G. Camara, and J.~A. Rosabal, \emph{{Flux algebra, Bianchi
  identities and Freed-Witten anomalies in F-theory compactifications}}, Nucl.
  Phys. \textbf{B814} (2009), 21--52,  \texttt{0811.2900}.

\bibitem{Font:2008vd}
A.~Font, A.~Guarino, and J.~M. Moreno, \emph{{Algebras and non-geometric flux
  vacua}}, JHEP \textbf{12} (2008), 050,  \texttt{0809.3748}.

\bibitem{Guarino:2008ik}
A.~Guarino and G.~J. Weatherill, \emph{{Non-geometric flux vacua, S-duality and
  algebraic geometry}}, JHEP \textbf{02} (2009), 042,  \texttt{0811.2190}.

\bibitem{Freed:1999vc}
D.~S. Freed and E.~Witten, \emph{{Anomalies in string theory with D-branes}},
  (1999),  \texttt{hep-th/9907189}.

\bibitem{Maldacena:2001xj}
J.~M. Maldacena, G.~W. Moore, and N.~Seiberg, \emph{{D-brane instantons and
  K-theory charges}}, JHEP \textbf{11} (2001), 062,  \texttt{hep-th/0108100}.

\bibitem{Evslin:2007ti}
J.~Evslin and L.~Martucci, \emph{{D-brane networks in flux vacua, generalized
  cycles and calibrations}}, JHEP \textbf{07} (2007), 040,
  \texttt{hep-th/0703129}.

\bibitem{Collinucci:2006ug}
A.~Collinucci and J.~Evslin, \emph{{Twisted homology}}, JHEP \textbf{03}
  (2007), 058,  \texttt{hep-th/0611218}.

\bibitem{Diaconescu:2000wy}
D.-E. Diaconescu, G.~W. Moore, and E.~Witten, \emph{{E(8) gauge theory, and a
  derivation of K-theory from M- theory}}, Adv. Theor. Math. Phys. \textbf{6}
  (2003), 1031--1134,  \texttt{hep-th/0005090}.

\bibitem{Kachru:2002gs}
S.~Kachru, J.~Pearson, and H.~L. Verlinde, \emph{{Brane/Flux Annihilation and
  the String Dual of a Non- Supersymmetric Field Theory}}, JHEP \textbf{06}
  (2002), 021,  \texttt{hep-th/0112197}.

\bibitem{Marchesano:2006ns}
F.~Marchesano, \emph{{D6-branes and torsion}}, JHEP \textbf{05} (2006), 019,
  \texttt{hep-th/0603210}.

\bibitem{Kaloper:1999yr}
N.~Kaloper and R.~C. Myers, \emph{{The O(dd) story of massive supergravity}},
  JHEP \textbf{05} (1999), 010,  \texttt{hep-th/9901045}.

\bibitem{Kachru:2002sk}
S.~Kachru, M.~B. Schulz, P.~K. Tripathy, and S.~P. Trivedi, \emph{{New
  supersymmetric string compactifications}}, JHEP \textbf{03} (2003), 061,
  \texttt{hep-th/0211182}.

\bibitem{Camara:2005dc}
P.~G. Camara, A.~Font, and L.~E. Ibanez, \emph{{Fluxes, moduli fixing and
  MSSM-like vacua in a simple IIA orientifold}}, JHEP \textbf{09} (2005), 013,
  \texttt{hep-th/0506066}.

\bibitem{Frey:2002hf}
A.~R. Frey and J.~Polchinski, \emph{{N = 3 warped compactifications}}, Phys.
  Rev. \textbf{D65} (2002), 126009,  \texttt{hep-th/0201029}.

\bibitem{Gurrieri:2002wz}
S.~Gurrieri, J.~Louis, A.~Micu, and D.~Waldram, \emph{{Mirror symmetry in
  generalized Calabi-Yau compactifications}}, Nucl. Phys. \textbf{B654} (2003),
  61--113,  \texttt{hep-th/0211102}.

\bibitem{Gurrieri:2002iw}
S.~Gurrieri and A.~Micu, \emph{{Type IIB theory on half-flat manifolds}},
  Class. Quant. Grav. \textbf{20} (2003), 2181--2192,  \texttt{hep-th/0212278}.

\bibitem{LoaizaBrito:2006se}
O.~Loaiza-Brito, \emph{{Freed-Witten anomaly in general flux
  compactification}}, Phys. Rev. \textbf{D76} (2007), 106015,
  \texttt{hep-th/0612088}.

\bibitem{Shelton:2006fd}
J.~Shelton, W.~Taylor, and B.~Wecht, \emph{{Generalized flux vacua}}, JHEP
  \textbf{02} (2007), 095,  \texttt{hep-th/0607015}.

\bibitem{Aldazabal:2006up}
G.~Aldazabal, P.~G. Camara, A.~Font, and L.~E. Ibanez, \emph{{More dual fluxes
  and moduli fixing}}, JHEP \textbf{05} (2006), 070,  \texttt{hep-th/0602089}.

\bibitem{Lawrence:2006ma}
A.~Lawrence, M.~B. Schulz, and B.~Wecht, \emph{{D-branes in nongeometric
  backgrounds}}, JHEP \textbf{07} (2006), 038,  \texttt{hep-th/0602025}.

\bibitem{Albertsson:2008gq}
C.~Albertsson, T.~Kimura, and R.~A. Reid-Edwards, \emph{{D-branes and doubled
  geometry}}, JHEP \textbf{04} (2009), 113,  \texttt{0806.1783}.

\bibitem{Hull:2004in}
C.~M. Hull, \emph{{A geometry for non-geometric string backgrounds}}, JHEP
  \textbf{10} (2005), 065,  \texttt{hep-th/0406102}.

\bibitem{Hull:2005hk}
C.~M. Hull and R.~A. Reid-Edwards, \emph{{Flux compactifications of string
  theory on twisted tori}},  (2005),  \texttt{hep-th/0503114}.

\bibitem{Hull:2009sg}
C.~M. Hull and R.~A. Reid-Edwards, \emph{{Non-geometric backgrounds, doubled
  geometry and generalised T-duality}},  (2009),  \texttt{0902.4032}.

\bibitem{Gukov:1999ya}
S.~Gukov, C.~Vafa, and E.~Witten, \emph{{CFT's from Calabi-Yau four-folds}},
  Nucl. Phys. \textbf{B584} (2000), 69--108,  \texttt{hep-th/9906070}.

\bibitem{Kachru:2002he}
S.~Kachru, M.~B. Schulz, and S.~Trivedi, \emph{{Moduli stabilization from
  fluxes in a simple IIB orientifold}}, JHEP \textbf{10} (2003), 007,
  \texttt{hep-th/0201028}.

\bibitem{deAlwis:2006cb}
S.~P. de~Alwis, \emph{{Transitions between flux vacua}}, Phys. Rev.
  \textbf{D74} (2006), 126010,  \texttt{hep-th/0605184}.

\bibitem{Blumenhagen:2009qh}
R.~Blumenhagen, M.~Cvetic, S.~Kachru, and T.~Weigand, \emph{{D-brane Instantons
  in Type II String Theory}},  (2009),  \texttt{0902.3251}.

\bibitem{Evslin:2003hd}
J.~Evslin, \emph{{Twisted K-theory from monodromies}}, JHEP \textbf{05} (2003),
  030,  \texttt{hep-th/0302081}.

\bibitem{LoaizaBrito:2003gz}
O.~Loaiza-Brito, \emph{{Instantonic branes, Atiyah-Hirzebruch spectral sequence
  and SL(2,Z) duality of N = 4 SYM}}, Nucl.Phys. \textbf{B680} (2004),
  271--301,  \texttt{hep-th/0311028}.

\bibitem{Evslin:2004vs}
J.~Evslin, \emph{{The cascade is a MMS instanton}},  (2004),
  \texttt{hep-th/0405210}.

\bibitem{Bouwknegt:2005ky}
P.~Bouwknegt, J.~Evslin, B.~Jurco, V.~Mathai, and H.~Sati, \emph{{Flux
  compactifications of projective spaces and the S- duality puzzle}}, Adv.
  Theor. Math. Phys. \textbf{10} (2006), 345--394,  \texttt{hep-th/0501110}.

\bibitem{Villadoro:2007yq}
G.~Villadoro and F.~Zwirner, \emph{{Beyond Twisted Tori}}, Phys. Lett.
  \textbf{B652} (2007), 118--123,  \texttt{0706.3049}.

\bibitem{Villadoro:2007tb}
G.~Villadoro and F.~Zwirner, \emph{{On general flux backgrounds with localized
  sources}}, JHEP \textbf{11} (2007), 082,  \texttt{0710.2551}.

\bibitem{Vafa:1996xn}
C.~Vafa, \emph{{Evidence for F-Theory}}, Nucl. Phys. \textbf{B469} (1996),
  403--418,  \texttt{hep-th/9602022}.

\bibitem{Douglas:1996du}
M.~R. Douglas and M.~Li, \emph{{D-Brane Realization of N=2 Super Yang-Mills
  Theory in Four Dimensions}},  (1996),  \texttt{hep-th/9604041}.

\bibitem{Sen:1996sk}
A.~Sen, \emph{{BPS states on a three brane probe}}, Phys. Rev. \textbf{D55}
  (1997), 2501--2503,  \texttt{hep-th/9608005}.

\bibitem{Hatcher:2002ht}
A.~Hatcher, \emph{{Algebraic Topology}}, Cambridge University Press. (2002),
  544 p.p.

\end{thebibliography}
\addcontentsline{toc}{section}{Bibliography}
\bibliographystyle{TitleAndArxiv} 
\end{document}